\documentclass[jgrga]{AGUTeX}

\usepackage[pdftex]{graphicx}
\usepackage{amsmath}
\usepackage{xfrac}
\usepackage{lineno}

\authorrunninghead{Fallows et al.}
\titlerunninghead{Ionospheric Scintillation Arcs}
\authoraddr{R. A. Fallows, ASTRON, 7990 AA Dwingeloo, the Netherlands (fallows@astron.nl) }

\begin{document}

\title{Broadband Meter-Wavelength Observations of Ionospheric Scintillation}

\author{R.A. Fallows,\altaffilmark{1} W.A.~Coles,\altaffilmark{2} D.~McKay-Bukowski,\altaffilmark{3,6} J.~Vierinen,\altaffilmark{3,5} I.I.~Virtanen,\altaffilmark{4} M.~Postila,\altaffilmark{3} Th.~Ulich,\altaffilmark{3} C-F.~Enell,\altaffilmark{3} A.~Kero,\altaffilmark{4} T.~Iinatti,\altaffilmark{3} M.~Lehtinen,\altaffilmark{3} M.~Orisp\"a\"a,\altaffilmark{4} T.~Raita,\altaffilmark{3} L.~Roininen,\altaffilmark{3} E.~Turunen,\altaffilmark{3} M.~Brentjens,\altaffilmark{1} N.~Ebbendorf,\altaffilmark{1} M.~Gerbers,\altaffilmark{1} T.~Grit,\altaffilmark{1} P.~Gruppen,\altaffilmark{1} H.~Meulman,\altaffilmark{1} M.~Norden,\altaffilmark{1} J-P.~de Reijer,\altaffilmark{1} A.~Schoenmakers,\altaffilmark{1} K.~Stuurwold,\altaffilmark{1} }

\altaffiltext{1}{ASTRON - the Netherlands Institute for Radio Astronomy, Postbus 2, 7990 AA Dwingeloo, the Netherlands}
\altaffiltext{2}{Electrical and Computer Engineering, University of California, San Diego, 9500 Gilman Drive, La Jolla, CA 92093-0403, USA}
\altaffiltext{3}{Sodankyl\"a Geophysical Observatory, University of Oulu, T\"ahtel\"antie 62, FI-99600 Sodankyl\"a, Finland}
\altaffiltext{4}{Department of Physics, P.O.Box 3000, FI-90014,
University of Oulu, Finland}
\altaffiltext{5}{MIT Haystack, 1 Millstone Hill Rd, Westford, MA 01886, United States}
\altaffiltext{6}{STFC Rutherford Appleton Laboratory, Harwell Science and Innovation Campus, Didcot OX11 0QX, UK}

\begin{abstract}
Intensity scintillations of cosmic radio sources are used to study astrophysical plasmas like the ionosphere, the solar wind, and the interstellar medium. Normally these observations are relatively narrow band. With Low Frequency Array (LOFAR) technology at the Kilpisj\"arvi Atmospheric Imaging Receiver Array (KAIRA) station in northern Finland we have observed scintillations over a 3 octave bandwidth. ``Parabolic arcs'', which were discovered in interstellar scintillations of pulsars, can provide precise estimates of the distance and velocity of the scattering plasma. Here we report the first observations of such arcs in the ionosphere and the first broad-band observations of arcs anywhere, raising hopes that study of the phenomenon may similarly improve the analysis of ionospheric scintillations. These observations were made of the strong natural radio source Cygnus-A and covered the entire 30-250\,MHz band of KAIRA. Well-defined parabolic arcs were seen early in the observations, before transit, and disappeared after transit although scintillations continued to be obvious during the entire observation. We show that this can be attributed to the structure of Cygnus-A. Initial results from modeling these scintillation arcs are consistent with simultaneous ionospheric soundings taken with other instruments, and indicate that scattering is most likely to be associated more with the topside ionosphere than the F-region peak altitude. Further modeling and possible extension to interferometric observations, using international LOFAR stations, are discussed. 
\end{abstract}

\begin{article}

\section{Introduction} 

The phenomenon of intensity scintillation (twinkling) of compact cosmic radio sources in the ionosphere was first identified in the 1950s and understanding it was one of the great successes of radio physics of that period.  It has also been observed in the solar wind and the interplanetary plasma where it has been equally useful.  The use of satellite beacons instead of cosmic sources has been very helpful in the ionospheric case particularly, especially as ionospheric scintillation can have serious consequences for our increasing reliance on satellite communication. This is not (yet) possible in the interstellar plasma, but recently radio astronomers discovered the ``parabolic arc'' phenomenon in dynamic spectral observations of interstellar scintillations of pulsars \citep{Stinebringetal:2001}.  These parabolic arcs are found in the two-dimensional (2-D) power spectrum of the observed dynamic spectrum. This 2-D power spectrum is referred to as the secondary spectrum. The parabolic arcs prove to be an ``eigenfunction'' of the broadband forward scattering problem and have revolutionized the study of the interstellar plasma. They have not proven to be as useful in the solar wind, although they have been observed (Coles, private communication), probably for two reasons. First, the major axis of the solar wind irregularities is normally aligned with the flow speed. Second, the inner scale of the solar wind turbulence is not very much smaller than the radius of the first Fresnel zone (as it is in the interstellar plasma). 

Ionospheric scintillation has also been probed for many years using radar measurements.  For this application various papers describe modelling calculations and results in which propagation through an irregular medium is detailed.  \citet{Nickisch:1992} and \citet{KneppandNickisch:2009}, for example, present phase screen/diffraction methods in which an electro-magnetic wave is propagated through multiple phase ``screens'' followed by a distance of free space to result in the calculation of the two-frequency, two-position mutual coherence function and its 2-D Fourier transform, termed variously as the ``scattering function'' or ``generalized power spectrum''.  These 2-D model spectra also show parabolic arc structures under some circumstances, most pronounced in the case of a single thin scattering screen as seen, for example, in Figure~ 11 of \citet{KneppandNickisch:2009}.  They are seen in radar observations, though examples reviewed in preparation of the current work do not show the phenomenon very clearly (e.g., Figure~ 1 in \citet{Nickisch:1992} showing results from an HF channel probe in northern Greenland and VHF results from an equatorial radar facility presented in \citet{Cannonetal:2006}).   

It has been unclear if parabolic arcs could be detected using observations of ionospheric scintillation of natural sources and we believe that the results presented in this paper represent the first (although \citet{KneppandNickisch:2009} note ``unpublished measurements of natural equatorial scintillation at 300\,MHz'' without making clear the parameters studied).  Here we show that they can be observed and have the potential to provide a 2-dimensional image of the scattering plasma as well as its velocity and distance.  

\begin{figure}
    \centering
    \includegraphics[width=8.2cm]{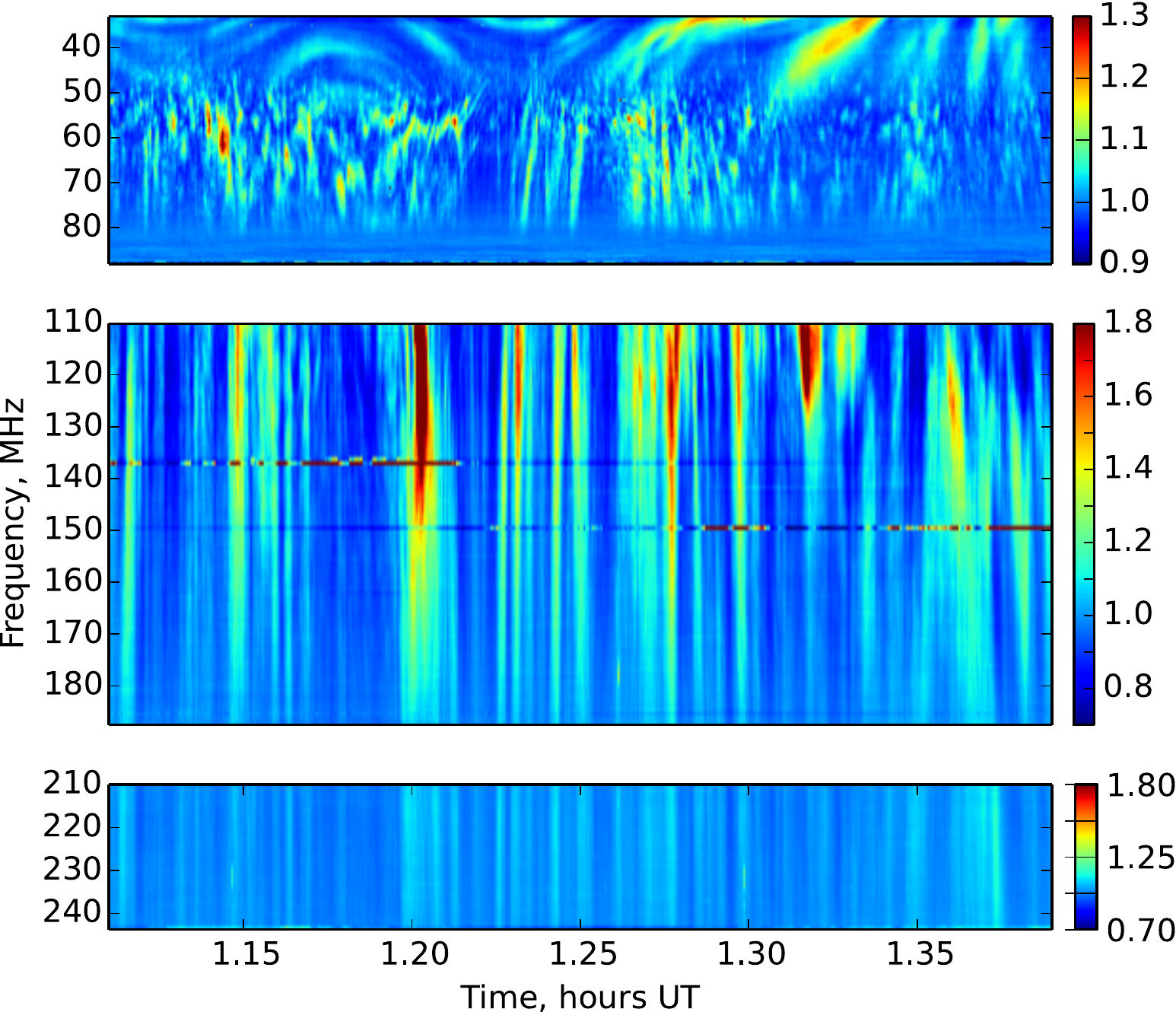}
    \caption{Dynamic spectra of a 17-minute segment of data from an observation of Cygnus-A on 25-26 September 2012. These spectra are simultaneous: the separation into different plots is to account for gaps in the frequency coverage and the differing scales used.  The raw intensity data for each frequency channel have been divided by their median to ``flatten'' the data for the frequency response across the pass-band.  The plot scale is linear with arbitrary intensity units.  }
    \label{fig:cygatwo}
\end{figure}

Improvements in radio technology have greatly increased the bandwidths available to radio astronomy and multi-GHz bands are now common, but multi-octave bands are not. The Low Frequency Array (LOFAR - \citet{LOFAR-reference-paper:2013}), a new radio telescope centered in the Netherlands with additional stations across Europe, has been the first to attempt to exploit these improvements at long wavelengths where they make possible a $3^+$-octave bandwidth. Observing over such bandwidths is qualitatively different than observing a few tiny windows over a 3-octave range. Long wavelength observations have become increasingly challenging due to the huge increase in wireless communications, but the resulting radio frequency interference (RFI) is much more easily identified (and subsequently removed) from a broad continuous band than from an array of small windows. This is most easily appreciated by examination of Figure~\ref{fig:cygatwo}, where some limited RFI is apparent at ~138 and 150\,MHz but this does not detract from the data in neighbouring channels. 

Broadband observations where dynamic spectra have been obtained are very few in number.  \citet{MeyerVernetetal:1981} observed ionospheric perturbations in decametric solar observations using the Nan\c{c}ay Decametric Array and interpreted their results in terms of diffraction and focussing by large-scale ionospheric disturbances.  \citet{Lecacheuxetal:2004} described wide bandwidth observations of various natural radio sources using the UTR-2 and URAN telescopes in Ukraine.  This included noting ionospheric scintillation in an observation of the strong radio source Taurus-A, but did not attempt any interpretation.

Here we describe the first wide-bandwidth ionospheric scintillation results from an observation of the strong radio galaxy Cygnus-A taken in September 2012 using the Kilpisj\"arvi Atmospheric Imaging Receiver Array - KAIRA \citep{KAIRA-reference-paper:2014}, a new ionospheric observatory built using LOFAR technology and summarised in Section~\ref{sec:KAIRA}.  These show remarkably complex refractive structures at the longest wavelengths, which are as yet unmodeled.  Secondary spectra (throughout this paper we will use the term ``secondary spectrum'' in this regard) of these data show parabolic arcs.  We also demonstrate here the use of straightforward scintillation arc modeling as used in the case of interstellar scintillation \citep{Cordesetal:2006}, which has the advantage of ease of use for the estimation of velocity and scattering height in the case of simple parabolic arcs.  Results from initial modeling attempts are presented and compared with available data from other systems at the time.  

Section~\ref{sec:arcs} describes the parabolic arc phenomenon and basic theory as it has been applied in the case of interstellar scintillation, with the observations and results presented in subsequent sections.

\section{Parabolic Arcs}
\label{sec:arcs}

The parabolic arc phenomenon is simply outlined, although the details are naturally more subtle.  The fundamental mechanism underlying intensity scintillation is the angular scattering of a plane wave incident on a thin scattering screen into an angular spectrum of plane waves $B(\theta)$. This angular spectrum is the distribution of power that one would observe with an imaging instrument in the receiving plane. Intensity fluctuations result from interference between these scattered plane waves. The basic observable is a dynamic spectrum of intensity $I(\nu,t)$, where $\nu$ is the observing frequency and $t$ is time through the observation, observed at a single point in the receiving plane as the spatial interference pattern drifts past the receiving antenna. The scattered plane waves each have a characteristic Doppler shift and delay which depend on the scattering angle ($\theta$), the wavelength of the radiation ($\lambda$), the distance ($L$) of the scattering screen, and its velocity ($V$) . The coordinates of the secondary spectrum, which is a delay-Doppler distribution, are actually differential delay and Doppler between pairs of scattered plane waves. In relatively weak scintillation the first order interference is between a scattered wave and the unscattered wave, so the secondary spectrum becomes a map of the scattered delay-Doppler distribution. A parabolic arc arises in the secondary spectrum naturally when one considers the delay and Doppler as a function of scattering angle. The delay is given by $\tau = (L/2c)|\theta|^2$ and the Doppler shift is $f= V \cdot \theta /\lambda$. Then $\tau$ is related to the maximum Doppler shift $f_{max}$ by $\tau = ( L \lambda^2 / 2 c V ^2 )  f_{max} ^2$ and the parabolic arc traces the boundary of the scattered distribution in the secondary spectrum. This boundary is quite sharp because the Jacobian of the transformation from $B(\theta)$ into the secondary spectrum has a half-order singularity at the arc.  In strong scattering, where interference of two scattered waves is dominant, the map is more complex and one frequently sees a field of reverse arclets with apexes along the primary arc. For detailed theory of arc formation in weak and strong scattering as it has been applied to interstellar scintillation the reader is referred to the work of \citet{Cordesetal:2006}.

The curvature of arcs defined in this way is $\lambda$ dependent, which is not important with 10\% bandwidths, but degrades the arcs significantly with octave bandwidths. Fortunately one can define the RF spectrum in $\lambda$ rather than $f$ and perform exactly the same analysis. The variable $\beta$, defined as the conjugate to $\lambda$, is $\beta = (\kappa/2\pi)^2 L/2$ where $\kappa$ is the spatial wavenumber of the scattering irregularities. In this form the Doppler shift is $f = V \cdot \kappa/2\pi$. Thus the arc is defined by $\beta = (L/2 V^2 ) f_{max}^2$ and the curvature is independent of $\lambda$.

\begin{figure*}
	\centering
    \includegraphics[width=5cm]{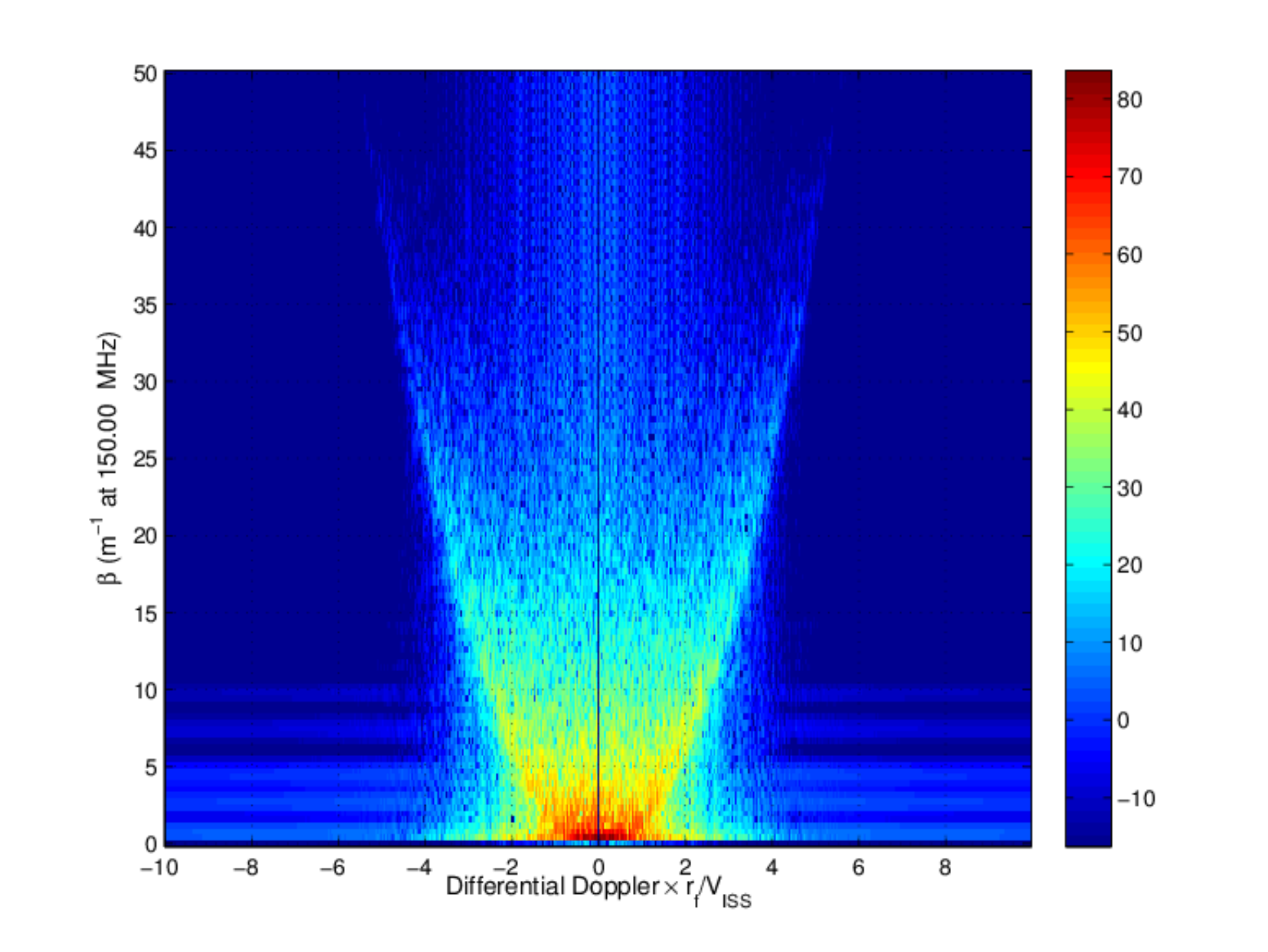} \includegraphics[width=5cm]{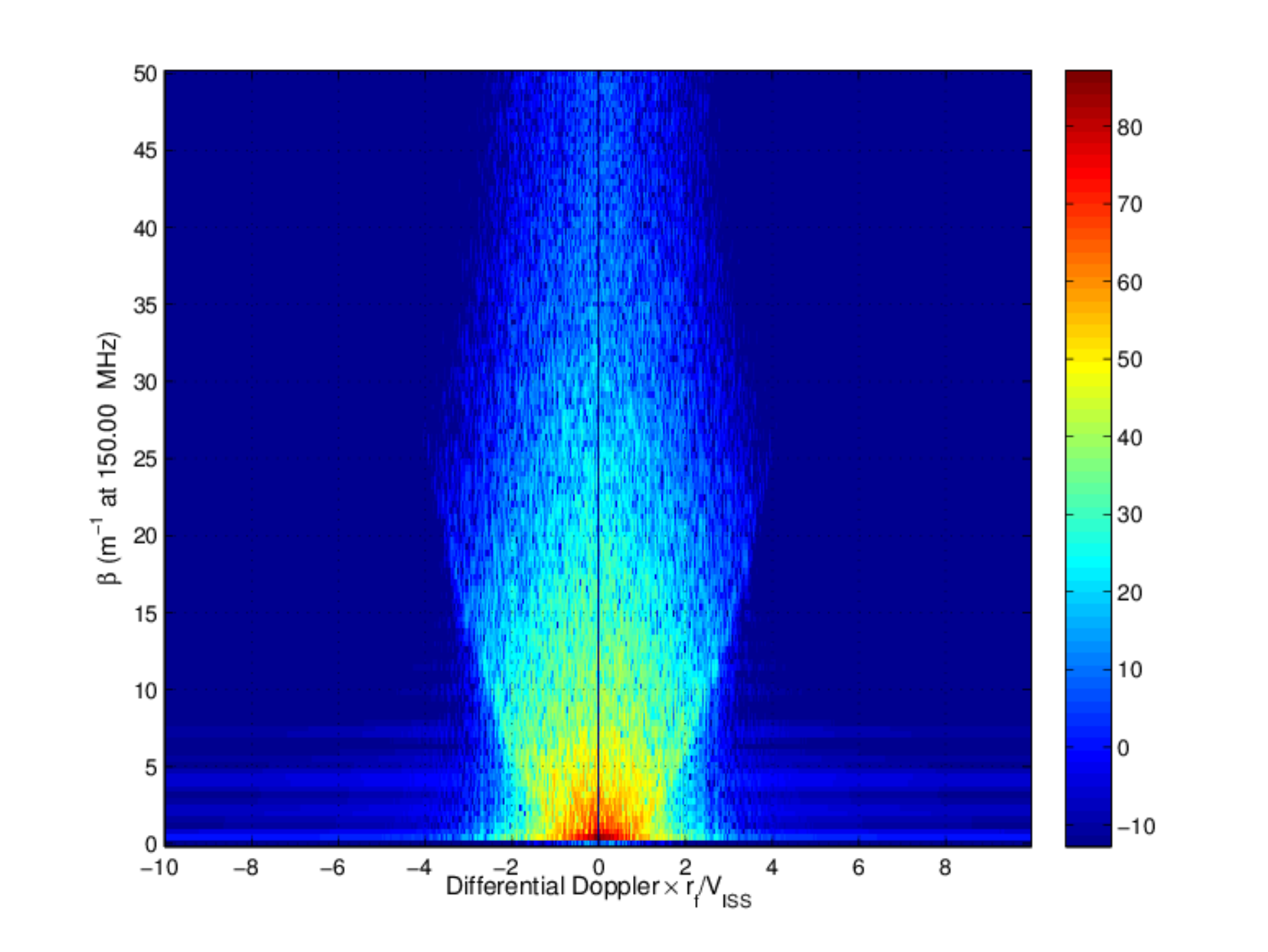} \includegraphics[width=5cm]{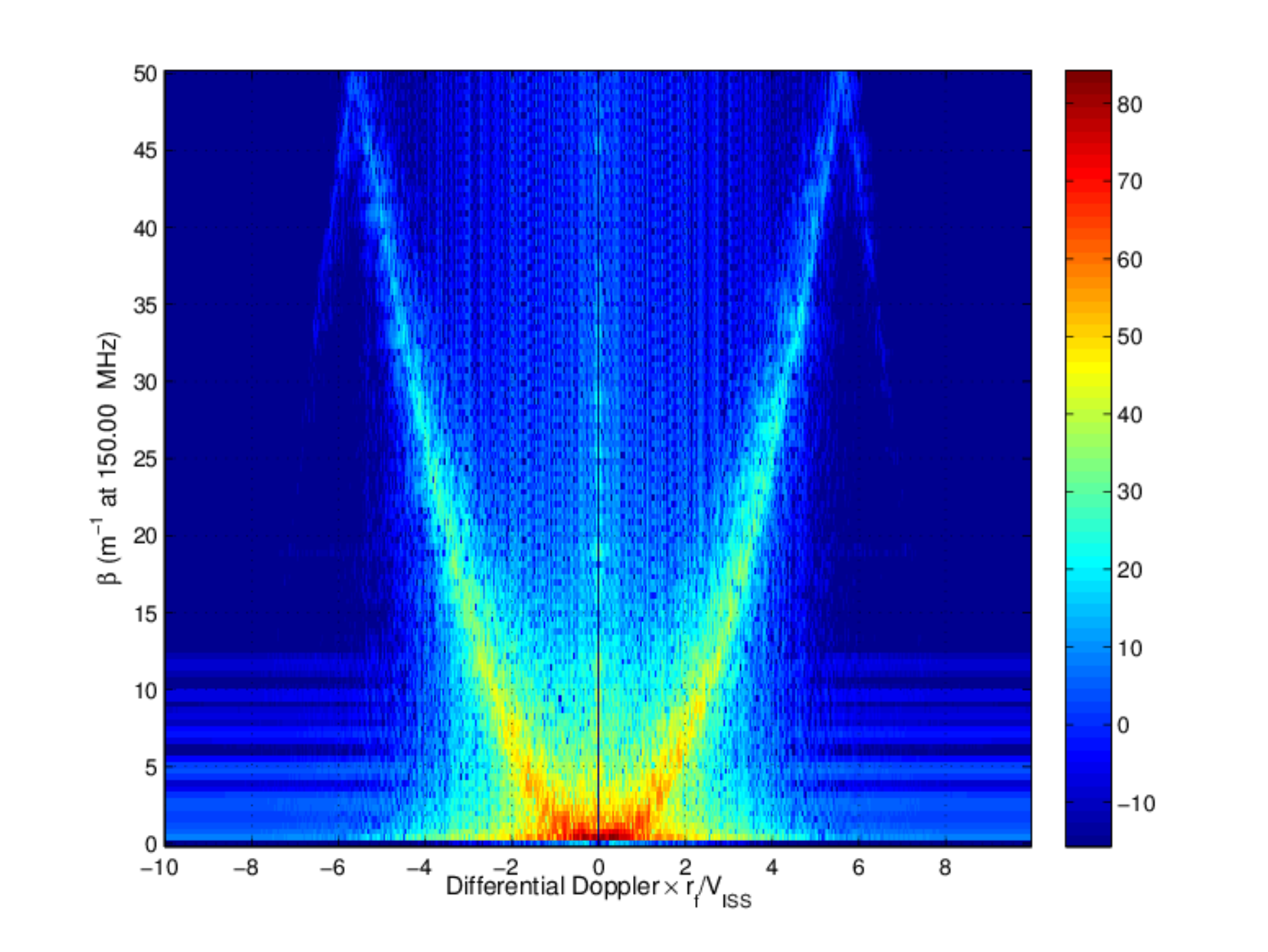}
    \caption{Secondary spectra of a 256\,s simulation.  The left panel is isotropic. The remaining panels have axial ratio 3:1: The middle panel shows the major spatial axis velocity aligned and the right is perpendicular.  Units are in dB.  }
    \label{fig:sim}
\end{figure*}

Parabolic arcs are much more distinct when $\theta \gg rms(\theta)$ so they are best observed with high signal to noise ratio and when $B(\theta)$ has a power-law tail. They are not visible if $B(\theta)$ is gaussian. The visibility is enhanced when $B(\theta)$ is anisotropic and $V$ is aligned with the more heavily scattered axis. This is most easily shown by a simulation. We have simulated 256\,s blocks of the ``High Band'' (Section~\ref{sec:KAIRA}) observations $\nu$ = 110 to 190\,MHz assuming the scattering is caused by Kolmogorov turbulence in the F region under three conditions: (1) the scattering is isotropic; (2) the axial ratio is 3:1 and the structures are aligned with the velocity; and (3) the axial ratio is 3:1 but the structures are aligned perpendicular to the velocity. We show the secondary spectra in Figure~\ref{fig:sim}.

In the solar wind the velocity is radial and the scattering is enhanced perpendicular to the magnetic field. Thus arcs are hard to see near the Sun in quiet conditions because the magnetic field is radial. They can be seen in coronal mass ejections where the magnetic field becomes non-radial. We expect that a similar anisotropy will hold in the ionosphere, but the velocity is more often perpendicular to the magnetic field. A concern in the ionosphere is that $B(\theta)$ may be truncated by an inner scale before $\theta \gg rms(\theta)$, eliminating the power-law tail. Conversely one potential application of this technique may be in estimating the dissipation scale of MHD turbulence in the ionosphere.

The effect of the angular size of the radio source $B_s(\theta)$ illuminating the ionosphere is also important. Effectively it smooths the spatial diffraction pattern $I(x,y)$ by convolution with a smoothing filter $S(x,y) = B_s(x=\theta_x/L, y=\theta_y/L)$. This multiplies the secondary spectrum by a low pass filter and truncates any parabolic arcs. The spatial scale of the diffraction pattern in weak scattering is $s_0 = \sqrt{\lambda L/2\pi}$ so the critical source diameter is $\theta_s < \sqrt{\lambda/(2 \pi L)}$. For our observations at $\lambda \sim 2\,m$ and $L \sim 400\,km$, the critical source diameter is 3 arc min. The radio source Cygnus-A is elongated $<$ 3 arc min in the minor axis and $>$ 3 arc min in the major axis. We believe this is a factor in our observations. An ideal source would be $<$ 3 arc min, but not too compact because if it is $<$ 0.3 arc sec it will scintillate in the solar wind and if it is $<$ 0.06 milli-arc sec it will scintillate in the interstellar plasma. Such scintillations would have to be separated from the ionospheric effects of interest. Choosing the right source diameter eliminates this confusion. As a practical matter only pulsars scintillate in the interstellar medium, but active galactic nuclei scintillate in the solar wind. Supernova remnants and compact radio galaxies are the best choice for ionospheric scintillations.

\section{Observations}

\subsection{Receiver System}
\label{sec:KAIRA}

KAIRA consists of a dual array of antennas: a Low-Band Antenna (LBA) array of 48 dual-polarisation dipole antennas which covers the frequency range 10--90\,MHz and a High-Band Antenna (HBA) array of 48 tiles, each tile consisting of a phased array of 16 dual-polarisation bow-tie antennas, which covers the range 110-270\,MHz.  The LBA dipoles and the HBA tiles are fixed in position, there is no mechanical beam steering.  The LBA dipoles are sensitive to the entire sky, although the sensitivity is significantly reduced below 25$^{\circ}$ elevation.  Each HBA tile has a 16 element analogue beam-former which creates a single tile-beam for each tile. The tile beam can be pointed anywhere inside the envelope of the dipole elements -- approximately the same ranges as the LBA dipoles. The analogue signals from each LBA dipole and each HBA tile (both in two polarizations) are carried to a receiver container on coaxial cables where they are amplified, filtered, digitized, separated into frequency channels, and full array-beams are formed for each channel. Multiple LBA array-beams can be pointed anywhere in the sky, but the HBA array beams are restricted to the field of view of the tile-beam, about 43$^\circ$ at 110\,MHz and 18$^\circ$ at 270\,MHz. The system is extremely flexible and can be operated in many different ``modes''.

The primary filters isolate the bands 10--90\,MHz, 110--190\,MHz and 210--270\,MHz. These bands are then channelized with a polyphase filter, producing 512~subbands each of width 195.3125\,kHz. The station beam-former is then used to form up to 244~``beamlets'', where each beamlet is an independent pointing direction and a specified subband for an arbitrary set of antennas from either of the two arrays. As the selection of antennas used can be specified separately for each beamlet, it is possible to perform an observation covering the full available bandwidth (10--270\,MHz) by using sub-arrays of LBA aerials and HBA tiles. However, given the limitation of a maximum of 244~beamlets available at the time, the frequency coverage is sparse as non-continuous subbands must be selected to span the entire frequency range.

For the observations presented here, the station was sub-arrayed into thirds to cover frequencies within each of the low-band and high-band filters used. In the low band 10--90\,MHz we selected 100 sub-bands (154--451 with spacing of 3) centered at 30.08 to 88.09\,MHz with spacing of 0.59\,MHz. In the high band 110-190\,MHz we selected 100 sub-bands (52--448, spacing 4) centered at 110.16 to 187.5\,MHz with spacing 0.78\,MHz. In the highest band 210--270\,MHz we selected 44 sub-bands (52--224, spacing 4) centered at 210.16 to 243.75\,MHz with spacing 0.78\,MHz.  Frequencies above 244\,MHz are heavily contaminated by RFI and so not regarded as useful in regular observations.

\subsection{Dynamic Spectra}
 
In September 2012, a series of test observations were carried out with the KAIRA station in which the strong radio source Cygnus-A was observed in overnight observations of, typically, 12 hours in length. Here we concentrate on the night of 25-26 September 2012. The 17 minute section shown in Figure~\ref{fig:cygatwo} is typical of the entire period. Since the scintillation time scales are typically 3 to 30 s, it is not possible to display the dynamic spectrum of the entire observation with enough resolution to see the structures. 

\begin{figure}
	\centering
    \includegraphics[width=8cm]{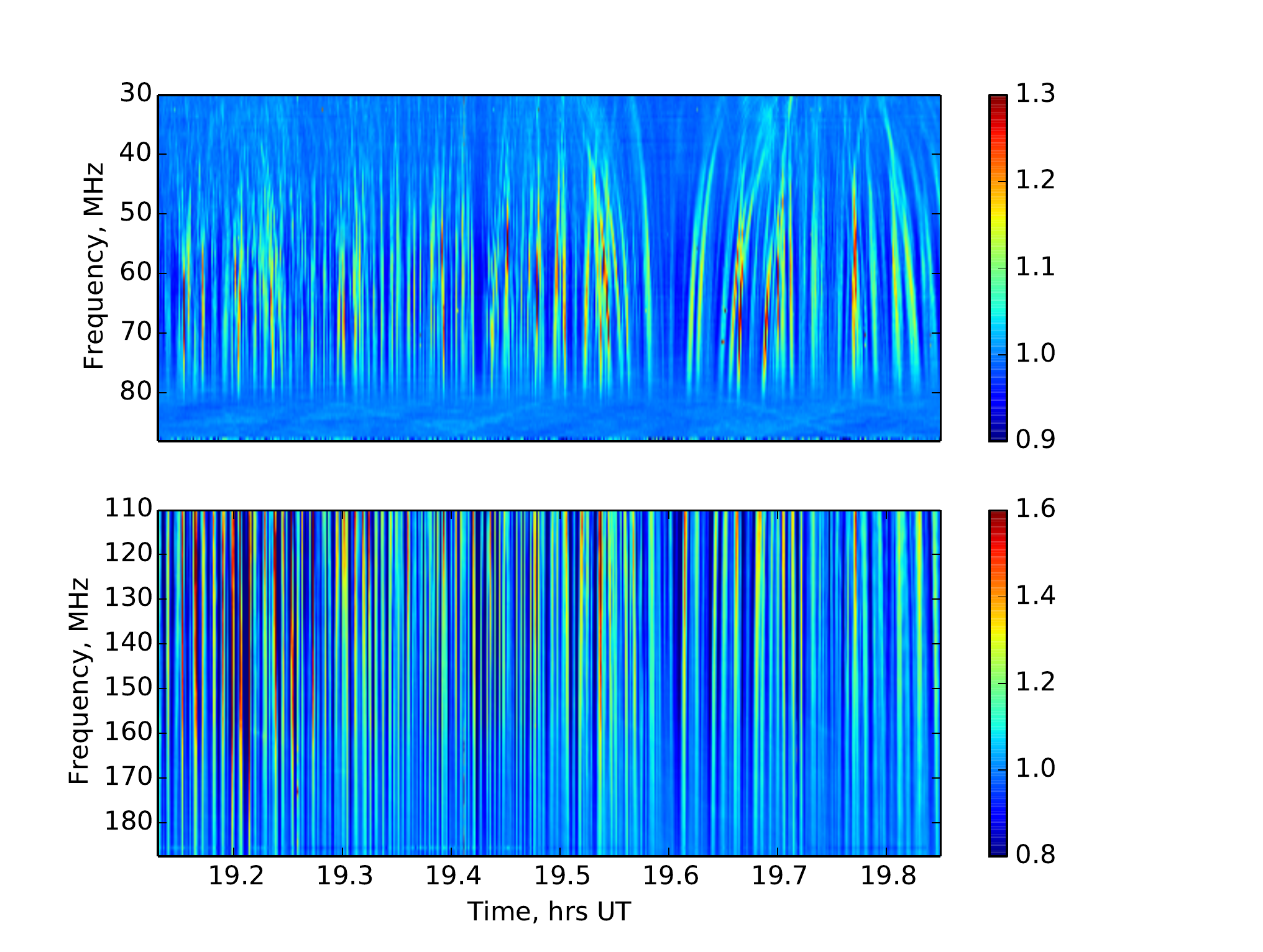}
    \caption{Dynamic spectra for the first 10 blocks of September 25-26 2012. The raw intensity data for each frequency channel have been divided by their median to ``flatten'' the data for the frequency response across the pass-band.  The plot scale is linear with arbitrary intensity units. }
    \label{fig:dyn1}
\end{figure}

\begin{figure}
	\centering
	\includegraphics[width=8cm]{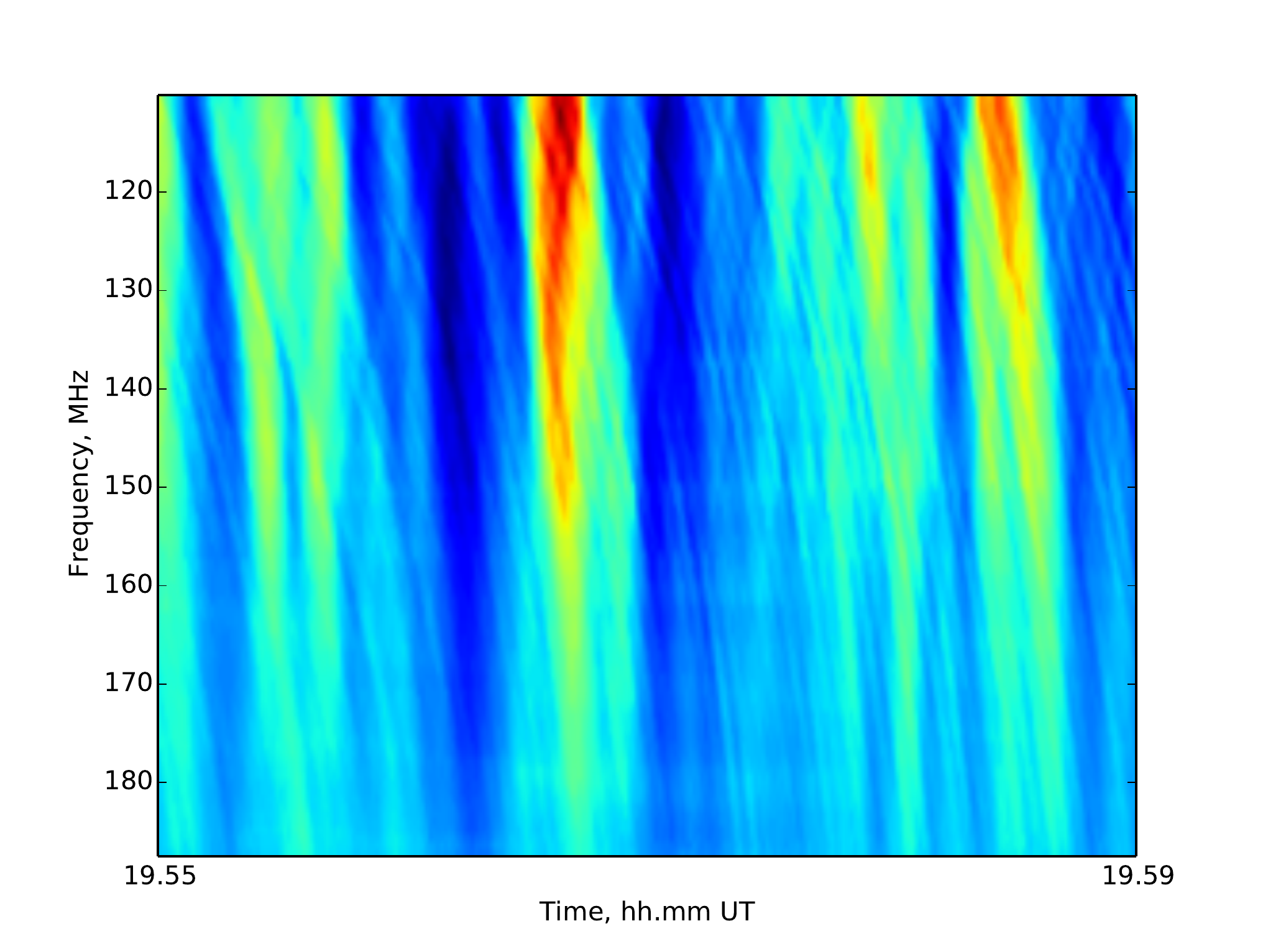}
	\caption{HBA dynamic spectrum for block 11 from the observation of 25-26 September 2012.  The raw intensity data for each frequency channel have been divided by their median to ``flatten'' the data for the frequency response across the pass-band.  The plot scale is linear with arbitrary intensity units. }
	\label{fig:dynzoom}
\end{figure}

The two higher bands show scintillations that are highly correlated over frequency as is characteristic of weak scattering.  The low band shows the onset of strong scattering and an obvious change in the lower part of this band to intensity structures with a much longer time scale and which decorrelate significantly in frequency.  Intensity structures with very short time scales are superposed on these long time scale structures.  This combination of time scales is indicative of a strong scattering regime where diffraction through small-scale density variations occurs alongside refraction through much larger-scale density structures which cause lensing effects leading to strong focussing and de-focussing of the radio source.  For further details the reader is referred to, e.g., \citet{Narayan:1992}. 

The dynamic spectra of the first 10 blocks of these two bands are shown in Figure~\ref{fig:dyn1}. The data in the low band above 80\,MHz is too heavily filtered to avoid contamination from the FM waveband to be of use so we truncated that band at 80\,MHz in computing secondary spectra. The low band spectra require more manual editing for RFI and we have not attempted to match them with the high band spectra because of the 30\,MHz gap. One can see that the scintillations are almost constant over the band although they vary rapidly with time. However there is fine structure on the scintillations, illustrated in Figure~\ref{fig:dynzoom}, and it is this fine structure that causes the parabolic arcs in the secondary spectra.

\begin{figure}
	\centering
    \includegraphics[width=7cm]{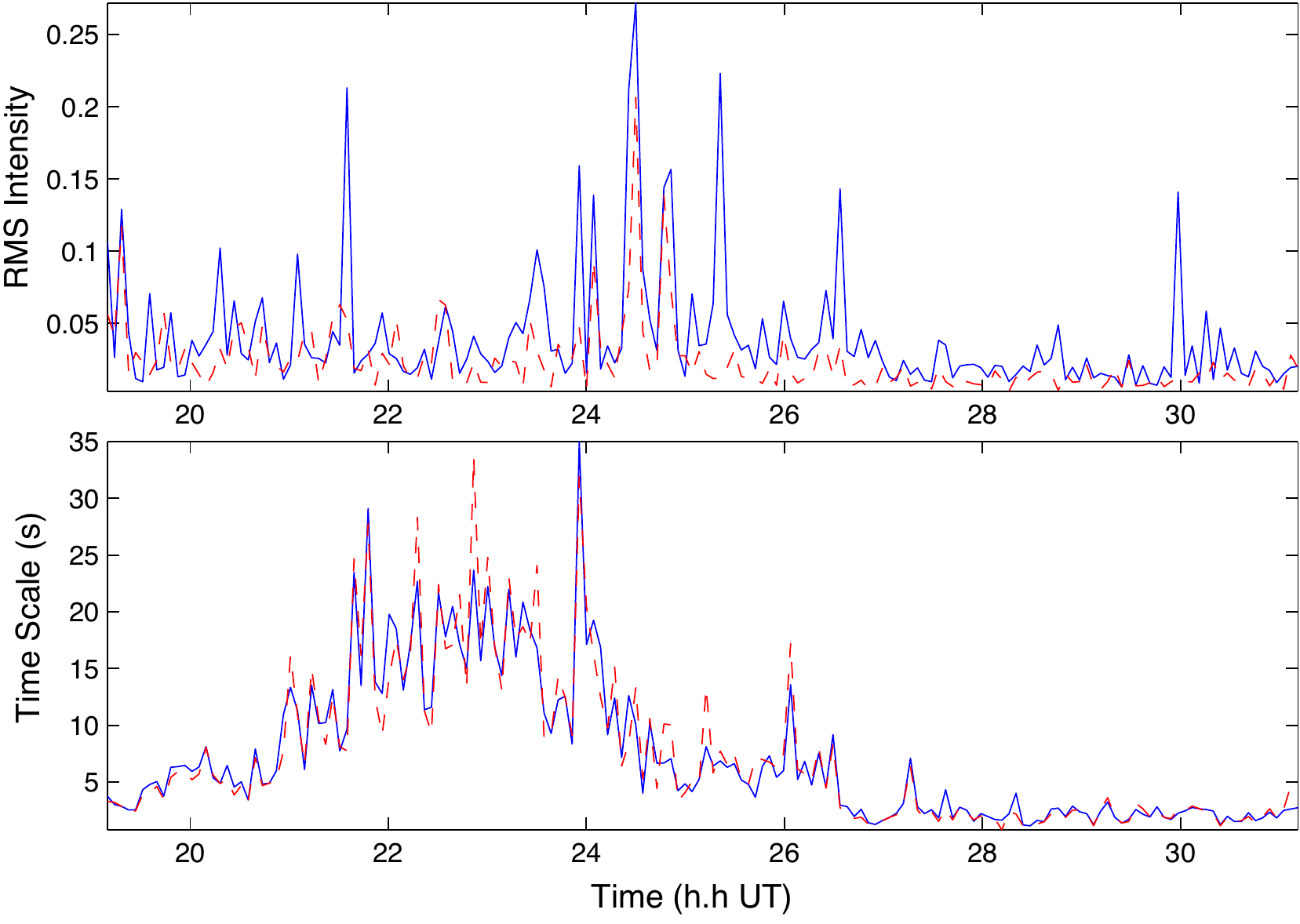}
    \caption{Scintillation index and time scale for the night of September 25-26 2012. The solid blue lines are from 110--149\,MHz and the dashed red lines are from 149--188\,MHz.}
    \label{fig:index_scale}
\end{figure}

We show $m = rms(I)/mean(I)$ and the time scale $\tau$, calculated as the mean period between intensity peaks in the scintillation pattern, from the mid-band in Figure~\ref{fig:index_scale}. The solid blue lines are from 110--149\,MHz and the dashed red lines are from 149--188\,MHz. One can see that both $m$ and $\tau$ are quite variable. We believe that the variations in $m$ reflect real variations in the level of turbulence, but that the spatial scale is relatively stable so the variations in $\tau$ reflect variations in velocity.  The scintillation index $m < 1$ throughout all three bands, but we believe that it would have exceeded unity in the low band were it not suppressed by the angular size of Cygnus-A.  A detailed analysis is beyond the scope of this paper because both the scattering and the source structure are anisotropic, and the source structure rotates with respect to the ionospheric structure during the observations, so a full 2-dimensional analysis is required.  This will be carried out in a future work.

\subsection{Secondary Spectra}

The secondary spectrum can be computed from the dynamic spectrum using a 2-dimensional Fourier transform
$|\tilde{I}(\tau,f)|^2 = |\mathcal{F}_2[I(\nu, t)]|^2$. Since the sampling in $(\nu,t)$ is regular an FFT can be used. However the spectrum is quite red and we find that best results are obtained by: pre-whitening \citep{JenkinsWatts:1969} the dynamic spectrum with a first difference on both axes; augmenting the pre-whitened dynamic spectrum with equal numbers of zero samples on both axes; performing the 2-D FFT; taking the squared magnitude; and finally post-darkening to correct for the pre-whitening. 

\begin{figure}
	\centering
    \includegraphics[width=4cm]{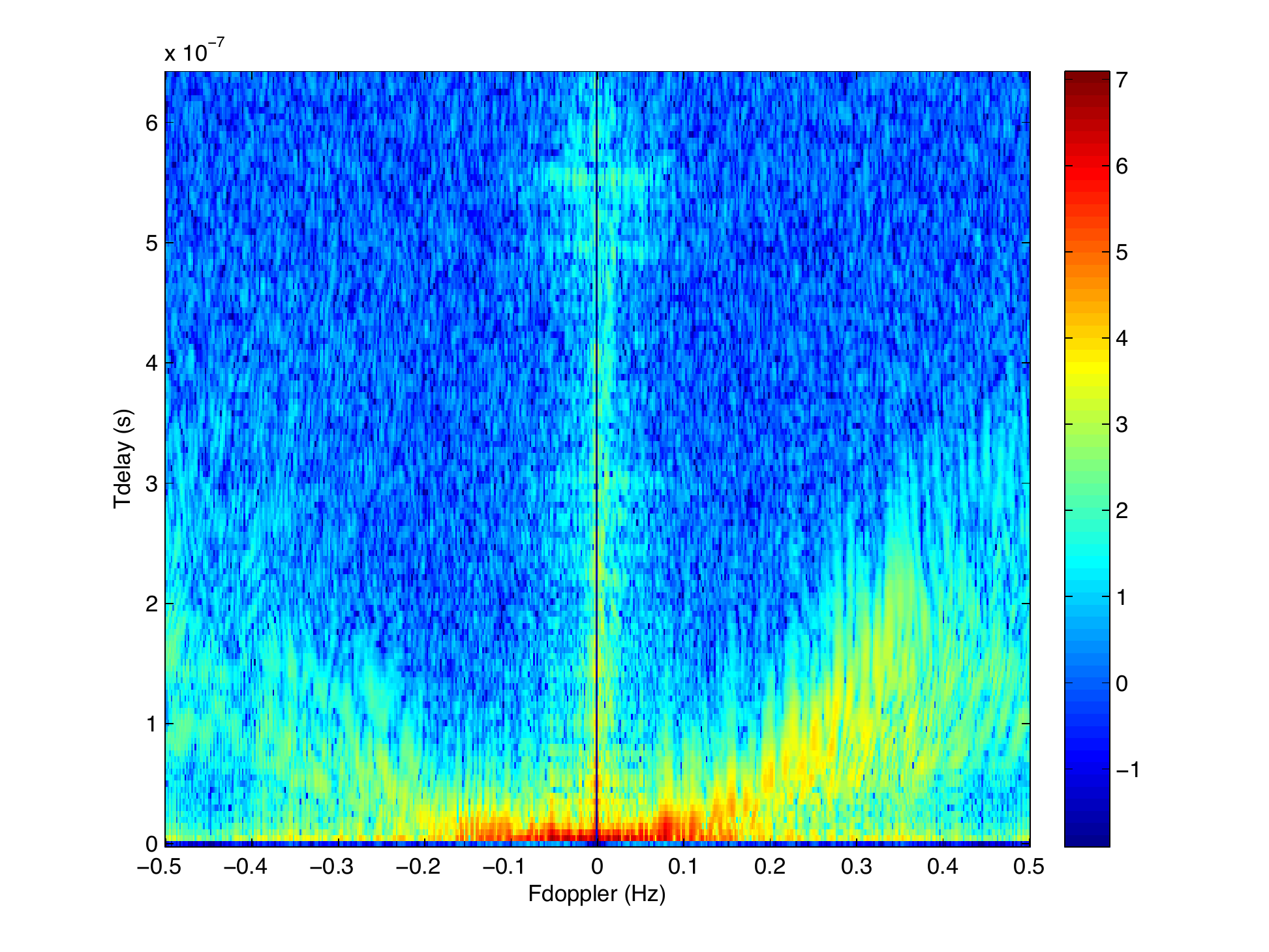} \includegraphics[width=4cm]{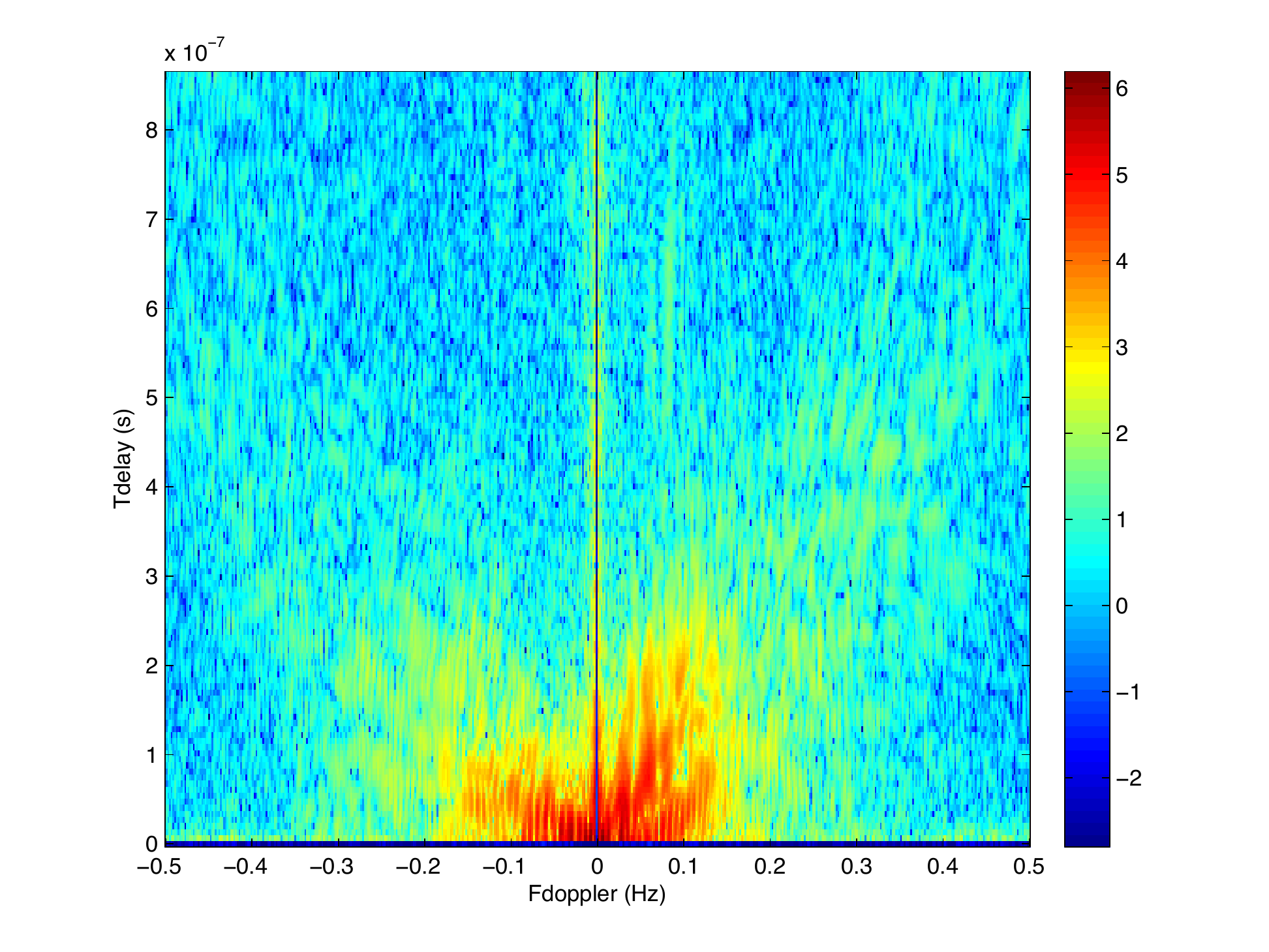} \\
     \includegraphics[width=4cm]{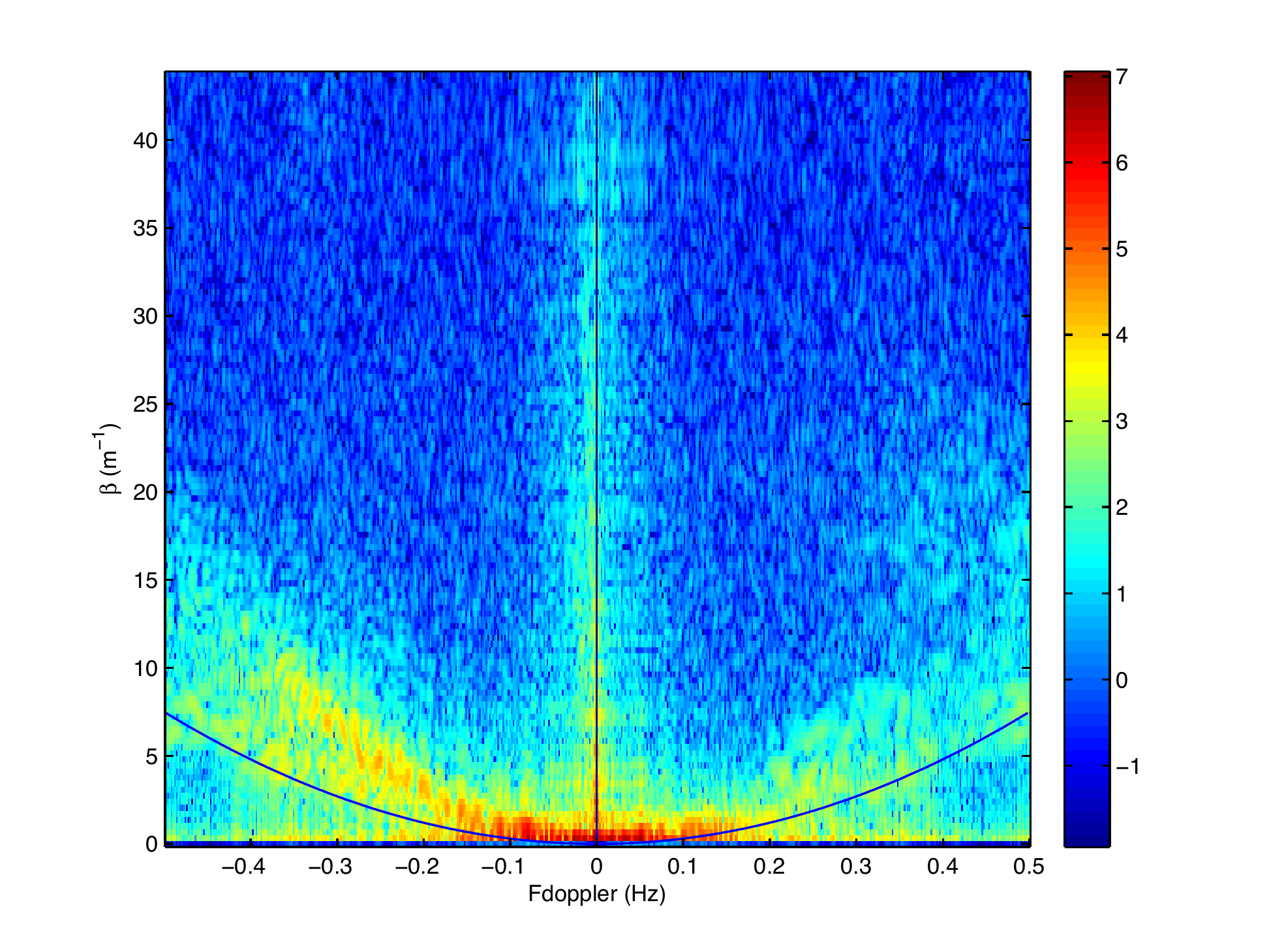} \includegraphics[width=4cm]{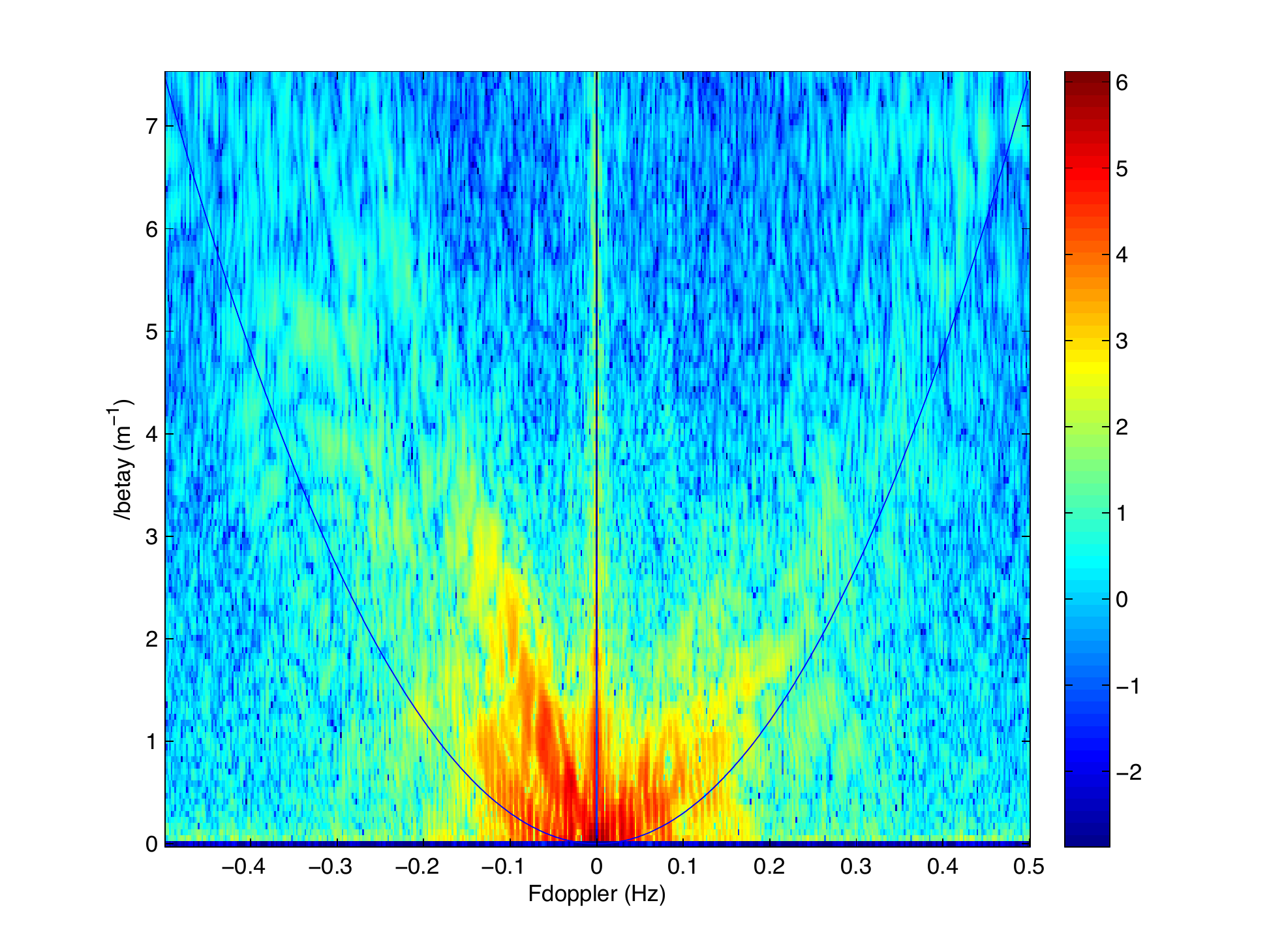} \\
    \caption{Secondary spectra of blocks 6 high band (left) and low band (right) of September 25-26.  The scale is logarithmic with units equivalent to dB/10. The top row is computed in equal $\nu$ steps and the bottom in equal $\lambda$ steps from the same set of data.  Note that switching between $\nu$ and $\lambda$ sampling reverses the Doppler axis.  These spectra show two interlaced arcs, most visible in the lower right plot.  The solid blue lines in the lower two plots are model fits to one of these arcs as detailed in section \ref{sec:arcfit}.  }
    \label{fig:chan}
\end{figure}

As noted earlier, with the octave bandwidth of these observations it would have been better to compute the spectra in equal wavelength channels. Since that cannot be done post-facto we have defined 100 equally spaced wavelength channels $\{\lambda_i \}$, calculated the corresponding frequency $\nu_i = c/\lambda_i$, and interpolated in the dynamic frequency spectrum to obtain $I(\lambda_i , t)$. We then compute the modified secondary spectrum $|\tilde{I}(\beta,f)|^2 = |\mathcal{F}_2[I(\lambda, t)]|^2$ using the FFT exactly as discussed earlier. Here $\beta$ is the conjugate variable to $\lambda$ and has units cycles/m. The effect of using equal $\lambda$ channels is easily seen in Figure~\ref{fig:chan}. This is a relatively complex secondary spectrum from the 6th block of Figure~\ref{fig:dyn1} which shows two interlaced parabolic arcs.  The top two plots in this figure show secondary spectra from high- and low-band parts of this block, and the lower two plots show secondary spectra from the same parts, but resampled to steps of equal $\lambda$.  The arcs are much better defined after resampling.  Here one should note that switching from equal $\nu$ to equal $\lambda$ sampling reverses the Doppler axis.

\begin{figure}
	\centering
    \includegraphics[width=8cm]{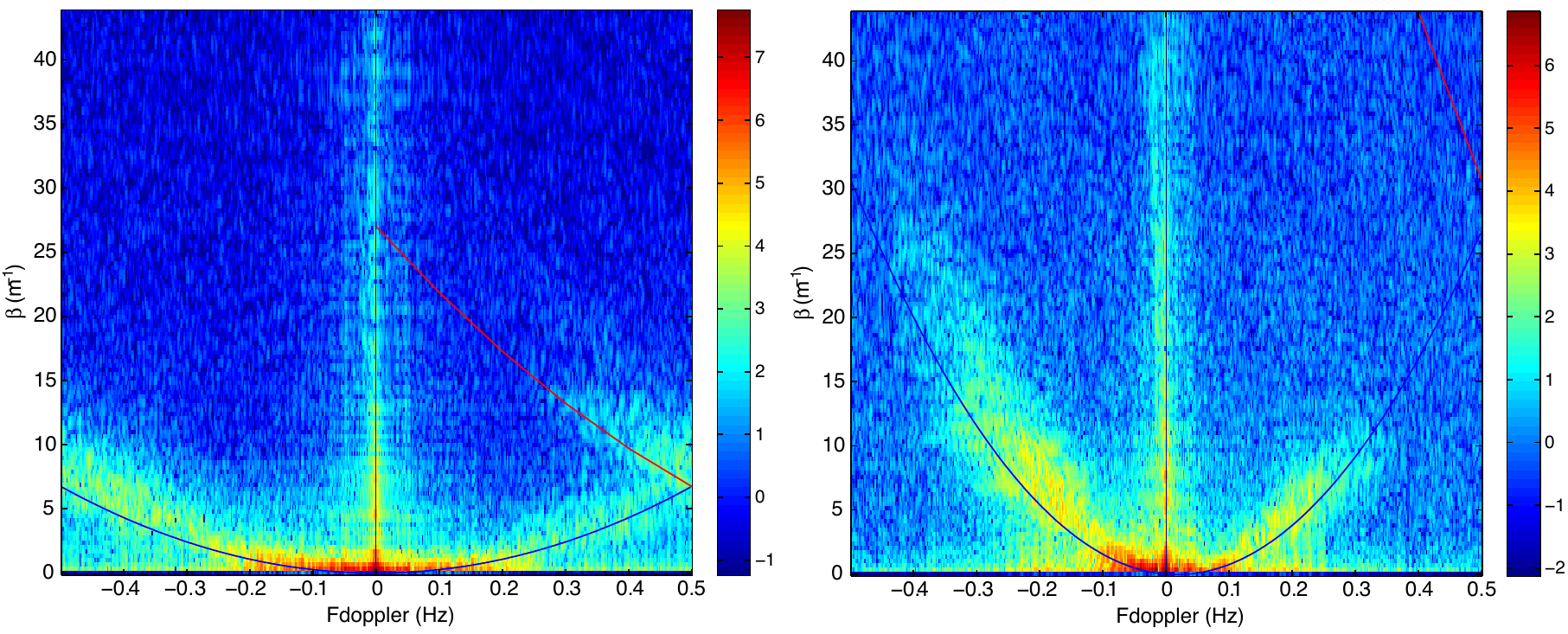}
    \caption{Secondary high band spectra of blocks 3 (left) and 10 (right) of September 25-26.  The scale is logarithmic with units equivalent to dB/10. The black curve is the main arc and the red one is the first alias.}
    \label{fig:sec1}
\end{figure}

The shortest sample interval that can easily be obtained with the KAIRA receivers is 1\,s.  Higher resolution is possible but requires the use of significant processing and data storage resources which were not readily available at the station at the time of these observations.  This is not always sufficient to capture the Doppler shifts, as can be seen in two examples; the high band secondary spectra for blocks 3 and 10 of Figure~\ref{fig:dyn1} which are shown in Figure~\ref{fig:sec1}. The left panel, block 3, has a relatively high velocity and the Doppler shift exceeds the Nyquist frequency at the right margin. The aliasing is clear because in 2-dimensions the delay (or $\beta$) continues to increase when the Doppler $f$ is aliased. In the right panel, block 10, the velocity is considerably lower and the entire parabola is well sampled.

Various time ranges of the dynamic spectra were used as input for calculating secondary spectra to investigate which ranges might be optimal; a time range of 256\,s represented a reasonable balance between ability to represent accurately the periodicities in the data and reflecting changes in these periodicities. We kept the high and low bands separate because the sampling rate was quite different, although they could have been analyzed together. We did not use the 44 channels of the top band because it would have left a large gap in the spectral range and the additional signal would not add any further information to the scintillation arc structure in the secondary spectra.

\begin{figure}
	\centering
    \includegraphics[width=8.1cm]{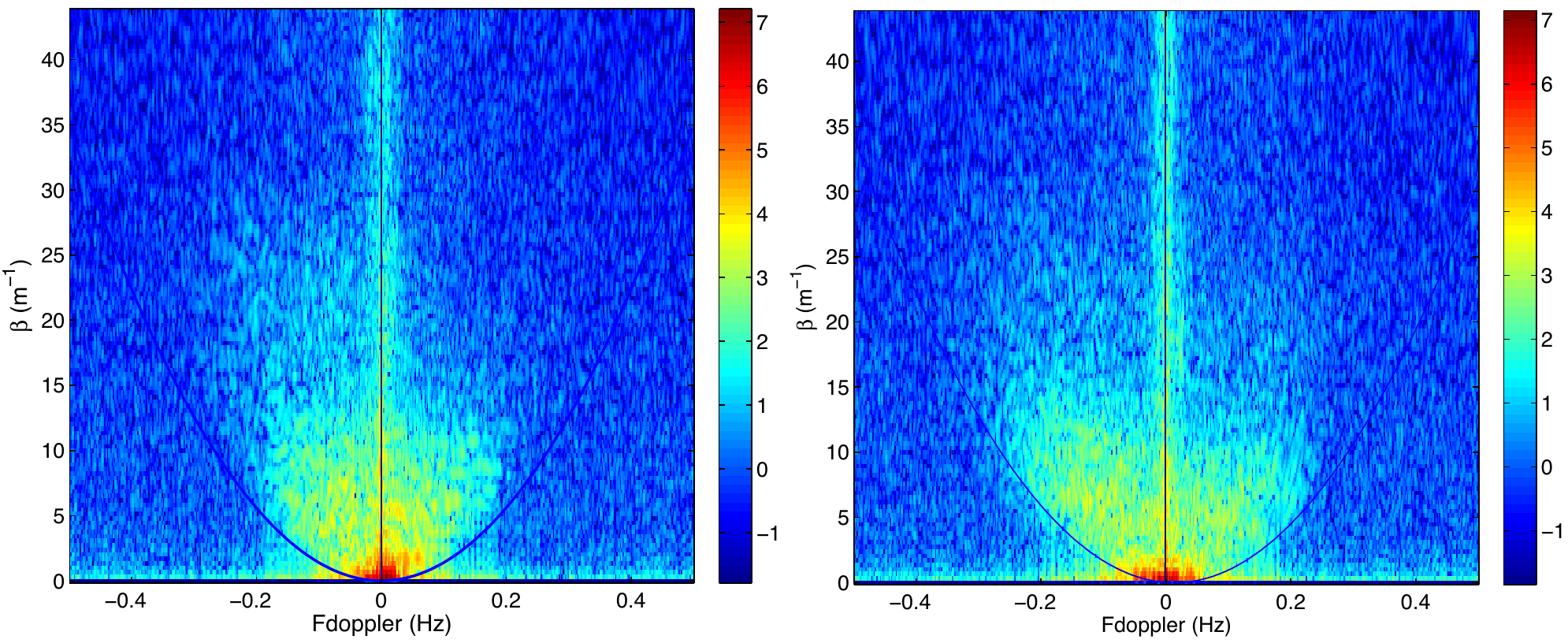}
    \caption{Secondary high band spectra of blocks 30 (left) and 52 (right) of September 25-26.  The scale is logarithmic with units equivalent to dB/10. Here the structure is isotropic or somewhat anisotropic but aligned with the velocity vector.}
    \label{fig:sec_iso}
\end{figure}

Parabolic arcs were observed in the first half of the data, but were much weaker in the second half and not observable after the time scale dropped at about 02:30 UT.  Note that the examples shown here have the character of the right panel of Figure~\ref{fig:sim}, i.e. the structure is elongated (presumably along the magnetic field) perpendicular to the velocity vector. This was the case for the early data, but after about 21:00 UT the secondary spectra look much more like the left panel of Figure~\ref{fig:sim}, or even a bit like the middle panel. The structure appears more isotropic or slightly anisotropic but aligned parallel with the velocity vector. Example blocks 30 and 52 at 21:12 and 22:48\,UT are shown in Figure~\ref{fig:sec_iso}. We believe that this is caused by the rotation of Cygnus-A with respect to the ionosphere: it is an elongated source whose long axis will have become more aligned with the local horizon as the source moved into the western sky through the latter part of the observation.  The source then begins to truncate the high angle scattering caused by the anisotropic microstructure making the scattering appear more isotropic.

\section{Analysis}

\subsection{Fitting the Parabolic Arcs}
\label{sec:arcfit}

The objective of this analysis is to show that we have detected parabolic arcs in ionospheric scintillation and that they are consistent with the other ionospheric measurements available. We will not attempt a complete model fitting at this time, however we note that model fitting in the weak scintillation case is much simpler than in the strong scintillation regime as shown in Appendix D of \citet{Cordesetal:2006}. Indeed one can further simplify the broad-band analysis by normalizing the dynamic spectrum $I(\lambda, t)$ vs $\lambda$, which one would normally do anyway because the radio source flux will vary with $\lambda$. In the normalized case the ``narrow band'' expression (D5) of \citet{Cordesetal:2006} applies even to a broad band. We also note that the effect of the finite size of the cosmic radio source illuminating the ionosphere will have to be included, as shown in their equation (D10). We have accurate radio images of Cygnus-A from interferometry so this correction is feasible.

In the following we will restrict our modeling to the parabolic arc itself, rather than to the entire secondary spectrum. We fit only the observations where we have complementary ionospheric observations and we fit only the blocks for which the best fit is obvious by eye. The arc should be a well-defined boundary between pure noise and scintillation, regardless of the axial ratio of the structure.  The arc defined by $\beta = (L/2 V^2 ) f_{max}^2$ has curvature $C = L/2 V^2$. We find that it is common to see an arc for which $\beta = C f_{max}^2 + B f_{max}$, a phenomenon which is caused by a large scale phase gradient which shifts the image of the radio source \citep{Cordesetal:2006}. We have included $B$ in the fit but we use only $C$ in the analysis. Examples are blocks 3 and 10 which are shown in Figure~\ref{fig:sec1}.

\begin{figure}[h]
	\centering
    \includegraphics[width=8.2cm]{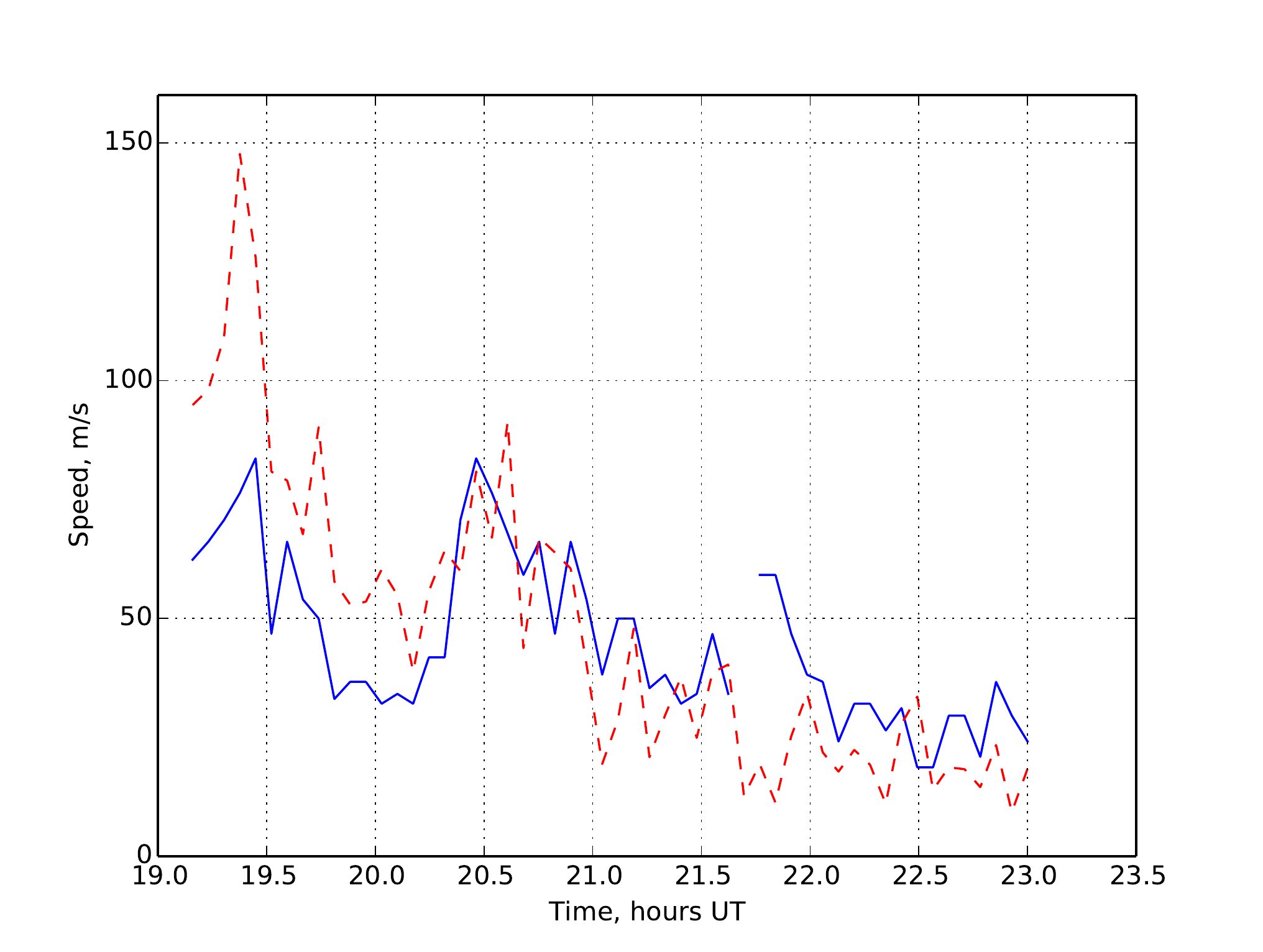}
    \caption{Velocity estimated from parabolic arcs (blue) and time scale (red). In both cases we assume the distance of the scattering region is 350 km.  The gap in the velocity time series is where a fit was not possible.}
    \label{fig:vel}
\end{figure}

The velocity can also be estimated from the time scale if the spatial structure is isotropic or if the major axis is perpendicular to the velocity. The latter case appears to be true up to about 23:00 UT and this is also the region for which it is easiest to estimate the location of the parabolic arc. We assume that the Fresnel scale is $r_f = (L \lambda/2\pi)^{0.5}$ and use estimated velocity $V_{est} = r_f /\tau$. We plot the velocity determined from the parabolic arcs, for cases where a fit could be made, and the spatial scales averaged over the high band in Figure~\ref{fig:vel}.  In both cases, we assume a distance along the line of sight to the scattering screen of 350\,km.

We choose the region between 19:46 and 20:12\,UT to compare with other ionospheric observations because the velocity is relatively stable. Block 10, which is shown in Figure~\ref{fig:sec1} is at the start of this interval and is typical of the other blocks.  Table~\ref{table:arcfits} gives the fitted values for blocks 10-16 corresponding to this time range.  The arcs seen in this time range are, however, relatively ``thick'', indicating scattering through a thick layer.  For some plots they appear almost to be visible as two separate arcs (as illustrated in Figure \ref{fig:chan}), an issue which is addressed in section \ref{sec:timeres}. The model fits given here represent average curvatures in such cases. 

\begin{table}
    \begin{tabular}{llll}
    Time & Block & Arc Curvatures & Velocity   \\
    hh:mm:ss UT & & Main & m\,s$^{-1}$ \\
    \hline
    19:46:44 & 10 & 160 & 33 \\
    19:51:00 & 11 & 130 & 37 \\
    19:55:16 & 12 & 150 & 34 \\
    19:59:32 & 13 & 170 & 32 \\
    20:03:48 & 14 & 150 & 34 \\
    20:08:04 & 15 & 170 & 32 \\
    20:12:20 & 16 & 100 & 42 \\
    \end{tabular}
    \caption{Table detailing the fitted curvature values for blocks 10-16 and the resulting calculation for the velocity, assuming a distance to the scattering screen of 350\,km.}
    \label{table:arcfits}
\end{table}

\begin{figure}[h]
	\centering
	\includegraphics[width=9cm]{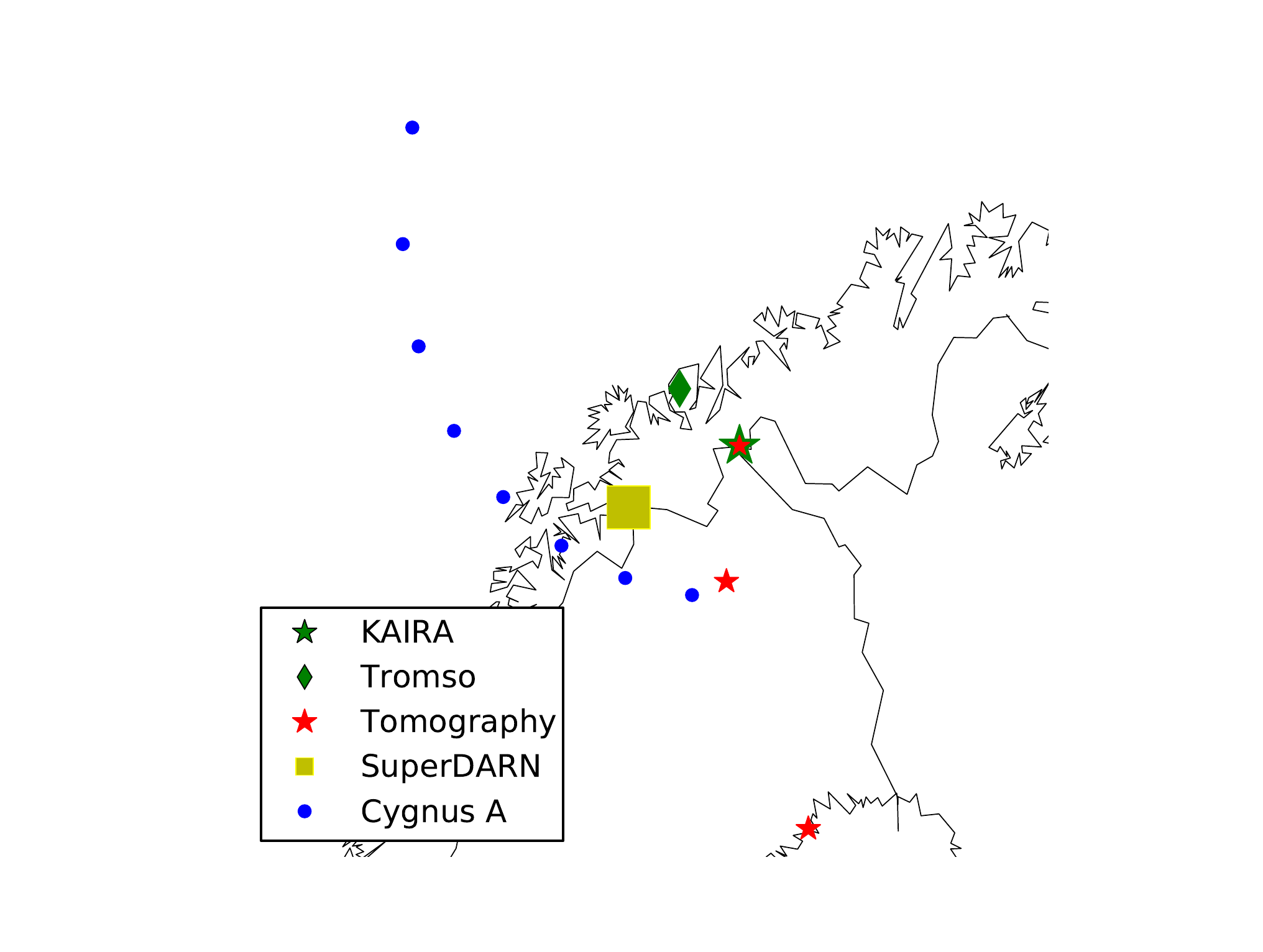}
	\caption{Map of northern Fenno-Scandinavia showing the location of points in the line of sight from KAIRA to Cygnus-A where it passes through a height of 300\,km (labelled as ``Cygnus-A''), for hourly times from 19:00\,UT (easternmost point) to 02:00\,UT (northernmost point) and the locations of KAIRA, the EISCAT site near Troms\o ~and Sodankyl\"a Geophysical Observatory ionospheric tomography receiver chain sites at (north to south) Kilpisj\"arvi, Kiruna and Lule\aa.  The approximate location of the SuperDARN measurements used is also indicated.}
	\label{fig:map}
\end{figure}

\subsection{Other Ionospheric Observations}

Any natural radio source is not stationary in the sky, but appears to move across the sky as the Earth rotates.  Cygnus-A is a circumpolar source at these latitudes.  Therefore to compare these data with other observations it is necessary to assess the approximate location of the ionospheric region giving rise to the scintillation.  To do this, we assume a peak F-region height of 300\,km and use this height to estimate the geographic location of a ``pierce point'' of the line of sight through the ionosphere.  Figure~\ref{fig:map} shows a map giving the locations of this point, calculated hourly for times from the start of the observation at 19:00\,UT (eastern-most point), in the line of sight (labelled as ``Cygnus-A'').  The location of the second point from the east corresponds to 20:00\,UT and, as the central time of the range for which we model the scintillation arcs, is the location used to compare with data from other observatories.  Unfortunately, there are no ionospheric observatories lying close enough to this point to provide a direct comparison, so only a general comparison can be made. 

Horizontal velocities in the ionosphere across the auroral regions are measured regularly by the Super Dual Auroral Radar Network (SuperDARN -- \citet{Chishametal:2007}, \citet{Greenwaldetal:1995}).  Since these radars measure Doppler velocities only in the direction of the radar beams, they are most useful if the fields of view of two or more radars cover the region to be probed.  The field of view of the Hankasalmi radar in Finland fully covers the area of ionosphere probed by the modelled scintillation measurements and the southernmost beam of the Pykkvibaer radar in Iceland covers an area down to approximately 0.5$^{\circ}$ in latitude to the north of the region of interest.  Taking this latter area, indicated on the map given in Figure \ref{fig:map}, velocity time series' have been obtained using the DaViT web tool for SuperDARN data currently hosted by VirginiaTech (http://vt.superdarn.org/tiki-index.php?page=DaViT+TSR).  Taking the closest area where a reasonable number of velocity points are available from both radars, the corresponding beams and range gates for each radar are beam 4, gate 13 of the Hankasalmi radar and beam 15, gate 36 of the Pykkvibaer radar.  Gaps in the time series' were interpolated across using a linear gradient between the measured velocities on either side of the gap. 
  
A correction was applied to the velocities to account for the estimated index of refraction, following the work of \citet{Gilliesetal:2009}.  The index of refraction, $n$, can be calculated using the plasma and radar frequencies, $f_{p}$ and $f$ respectively using the basic formula, $n = \sqrt{1-(f_{p}^{2}/f^{2})}$.  Information available using the range-time plotting function of the DaViT tool (http://vt.superdarn.org/tiki-index.php?page=Range+Time+Browse) indicates that both Hankasalmi and Pykkvibaer were operating at a frequency of 10\,MHz at the time.  The electron plasma frequency is determined from $f_{p} = 8.98n_{e}^{0.5}$ where $n_{e}$ is the electron density.  Using an F-region peak density of $2.3x10^{11}\,m^{-3}$ estimated from ionospheric tomography reconstructions (detailed below - see Figure \ref{fig:tomo}), $f_{p}$ is calculated to be 4.26\,MHz.  This leads to a refractive index of 0.9 for both radars and their velocities were multiplied by a factor of $1/n$.
  
\begin{figure}
	\centering
	\includegraphics[width=9cm]{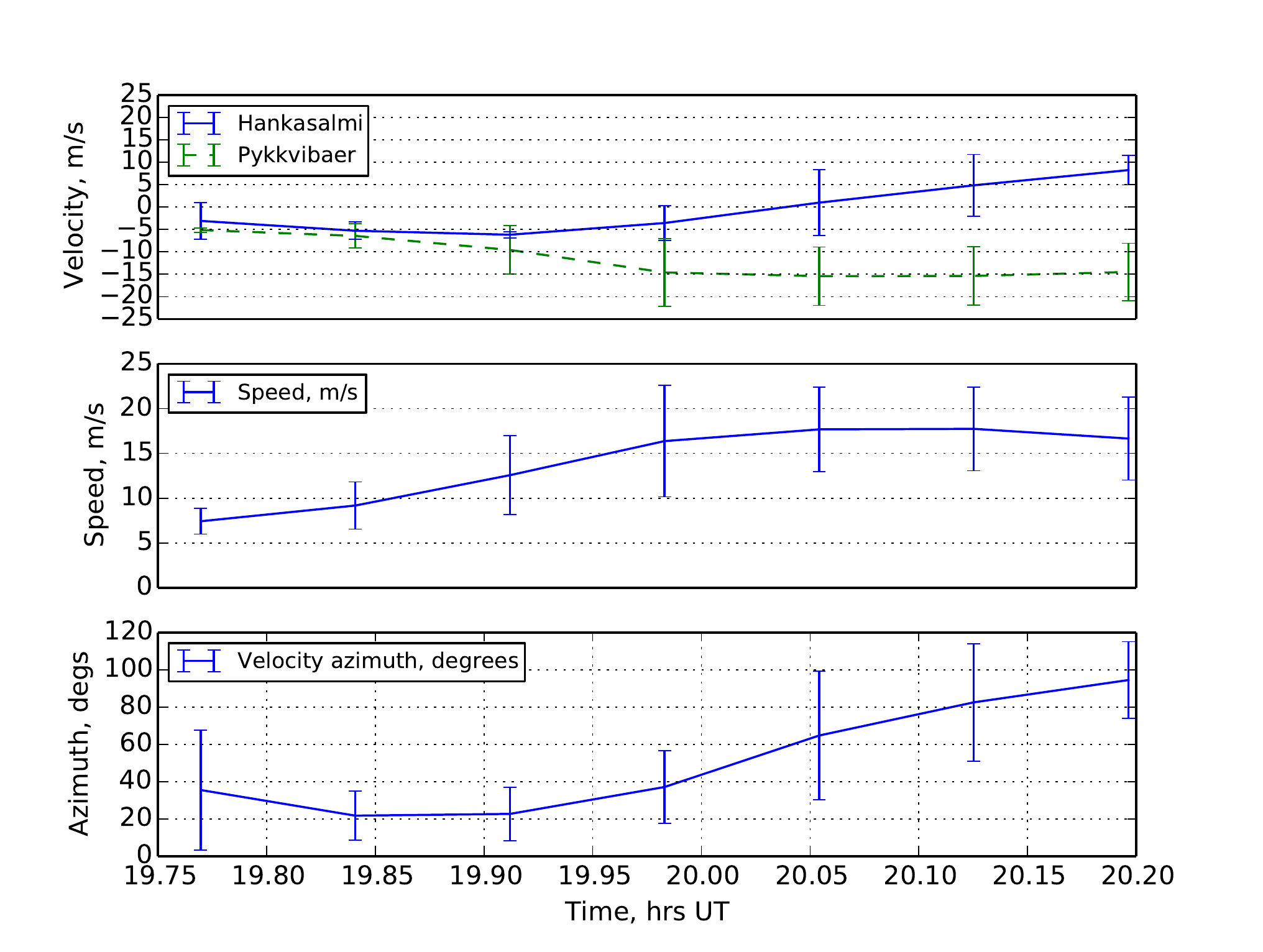}
	\caption{SuperDARN velocities for blocks 10-16.  These are calculated from averages of four data points covering the length of each block with the standard deviation taken to be the error.  The top plot shows the velocities from each radar separately; the middle plot shows the velocity magnitude after combining them; the lower plot shows the velocity direction in degrees azimuth measured eastwards from north.}
	\label{fig:superdarn}
\end{figure}

    The final vectors were averaged over each block (four data points per block) and the standard deviation taken to be the error.  The azimuths, measured eastwards from north, of the boresites of each of these beams are 336.66$^{\circ}$ for Hankasalmi and 57.54$^{\circ}$ for Pykkvibaer.  Using these angles the velocities from each radar were combined into velocity magnitude and direction.  The resulting velocities and directions are given in Figure \ref{fig:superdarn}.

\begin{figure}
	\centering
	\includegraphics[width=9cm]{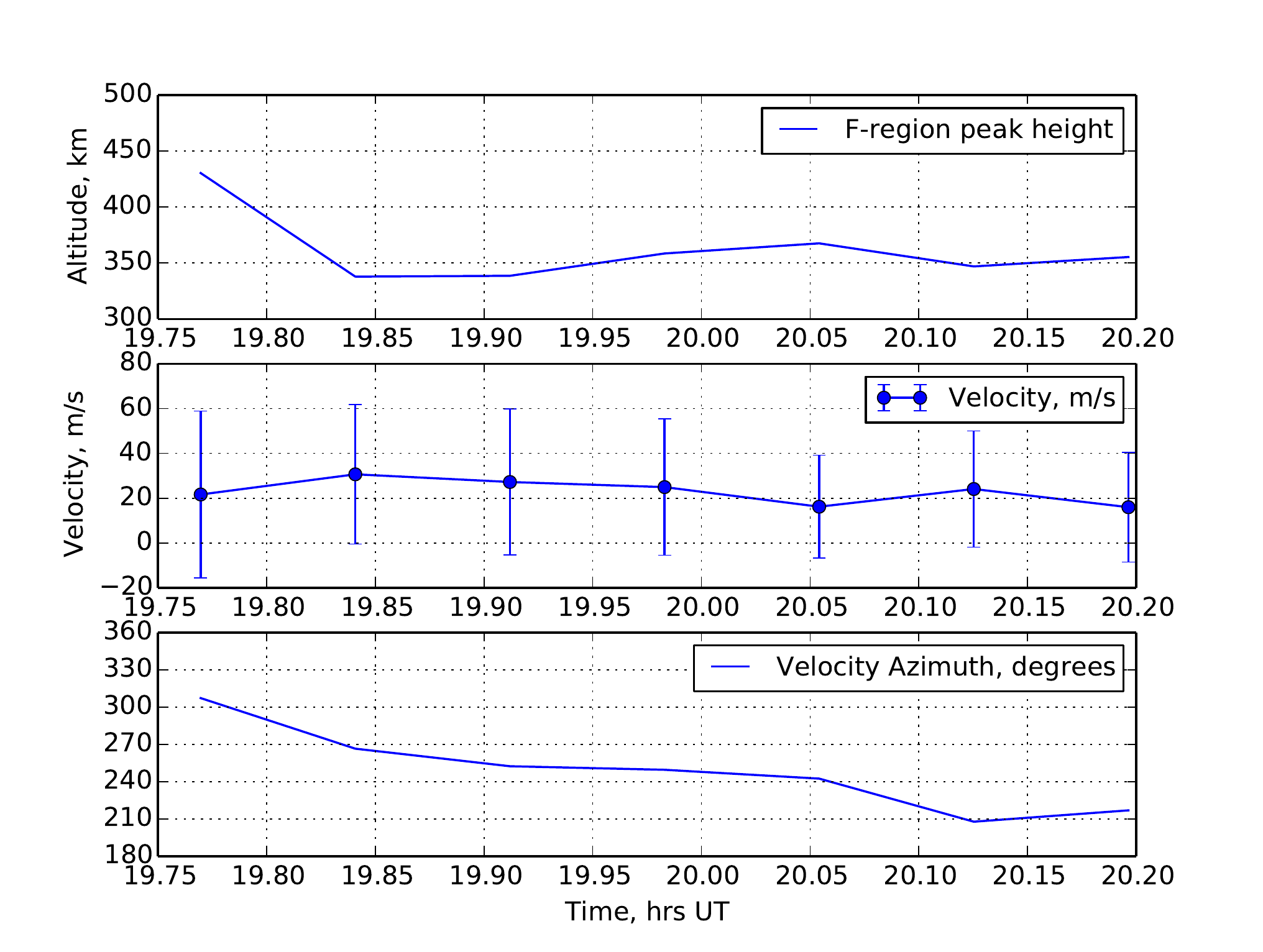}
	\caption{Troms\o~dynasonde data for the F-region peak height (top), F-region horizontal velocity magnitude (middle) and F-region horizontal velocity direction in degrees azimuth, measured eastwards from north (bottom) for blocks 10-16.}
	\label{fig:dynasonde}
\end{figure}

\begin{figure}
	\centering
	\includegraphics[width=9cm]{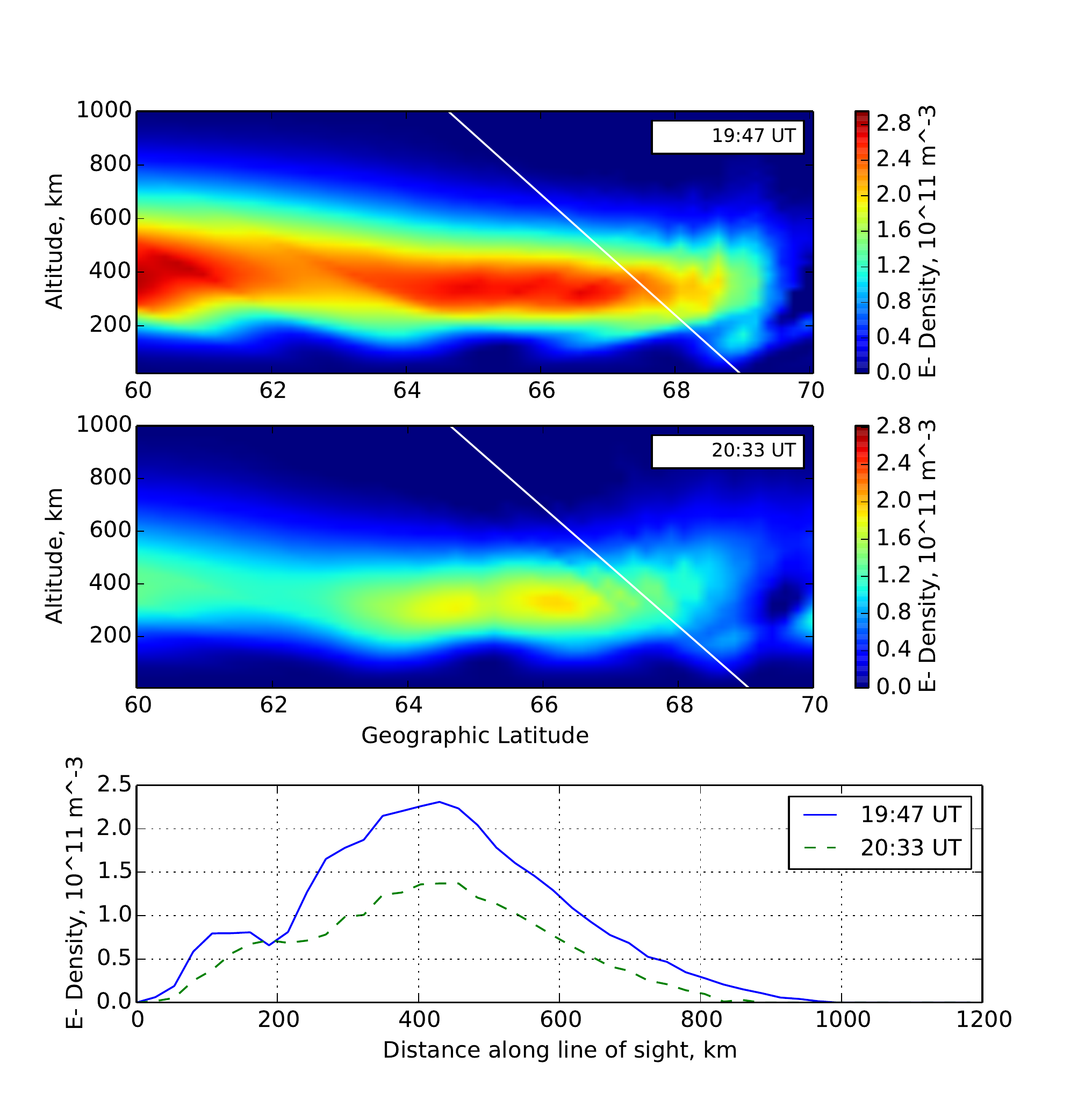}
	\caption{The top two plots are tomographic reconstructions, using the Chapman model, of satellite passes at 19:47\,UT and 20:33\,UT on 25 September 2012 recorded by the Sodanyl\"a Geophysical Observatory tomography chain.  The white line indicates the approximate line of sight from Kilpisj\"arvi towards Cygnus-A. Density profiles estimated from the reconstructed electron density values along the line of sight are shown in the lower plot for each satellite pass.}
	\label{fig:tomo}
\end{figure}

The Troms\o~dynasonde \citep{Rietveldetal:2008}, based at the European Incoherent Scatter (EISCAT, \citet{RishbethWilliams:1985}) radar site at Ramfjordmoen in Norway, produces a number of ionospheric parameters including velocity vectors.  Although it is not very close to the region of interest here (as shown in Figure~\ref{fig:map}), it is the closest such instrumentation which was operating at the time and gives an idea of the prevailing ionospheric conditions in the region.  Figure~\ref{fig:dynasonde} presents plots of the horizontal velocities in the F-region, and the F-layer peak height for the times of interest.  The dynasonde also produces data for the vertical velocity and similar data for the E-region.  The vertical F-region velocity was approximately zero at the time.

Further information on the density structure of the ionosphere at this time can be obtained from ionospheric tomography data.  Sodankyl\"a Geophysical Observatory operate a chain of ionospheric tomography receivers across Finland and Sweden, with the highest latitude station at Kilpisj\"arvi, close to KAIRA.  This chain recorded satellite passes at 19:47\,UT and 20:33\,UT on 25 September 2012, with  the satellites following paths a little to the west of the coast of Norway and over the Finland/Russia border respectively.  Plots of these tomographic reconstructions (the reader is referred to \citet{Nygrenetal:1997} for a description of the inversion process), using a Chapman function as
regularization profile \citep{Markkanenetal:1995}, are given in Figure \ref{fig:tomo}.  

The reconstructions indicate maximum density at approximately 400\,km and demonstrate a possible lowering of the overall electron density between 19:47 and 20:33\,UT, but otherwise no significant change in the overall structure of the ionosphere over this time frame.

\subsection{Estimating Scattering Height}
\label{sec:scatheight}

Given the external velocity measurements of SuperDARN and the Troms\o~dynasonde, it is possible to use the scintillation arc curvatures to estimate the height of the dominant scattering ``layer''.  However, the velocity which is used in modelling the scintillation arcs is the result of both movement of the ionosphere and movement through the ionosphere of the line of sight to the radio source, and is sensitive only to any direction perpendicular to the line of sight.  

It is first sensible to define co-ordinates in the reference frame of the line of sight:

\noindent $x$: Horizontal with respect to the local horizon, positive clockwise.

\noindent $y$: Vertical with respect to the local horizon, positive upwards.

\noindent $z$: Within the line of sight and therefore not used.

The velocity of the line of sight is itself naturally dependent on the distance used.  Calculations show that it is of the order of 25\,m\,s$^{-1}$ assuming a distance of 350\,km, so is comparable in magnitude to the velocities measured by the radar systems at the time and cannot be neglected.

We assume also that the velocity component due to the ionosphere is horizontal (i.e., parallel to the ground plane):  This is a reasonable assumption given that SuperDARN only measures horizontal velocities and the Troms\o~dynasonde data give no reliable indication of any significant vertical velocity component.  

The velocity of the line of sight can be stated in terms of an assumed distance along it and the angle through which the radio source moves within the time covered by each data block.  Using also the change in elevation of the radio source enables the velocity of the line of sight to be stated in terms of horizontal and vertical components, as given in Equations \ref{eqn:vsrchoriz} and \ref{eqn:vsrcvert}:

\begin{linenomath*}
\begin{equation}
	V_\mathrm{x,los} = L \times 2\sin(\sfrac{{\theta}_\mathrm{x}}{2})/t
	\label{eqn:vsrchoriz}
\end{equation}
\end{linenomath*}

\begin{linenomath*}
\begin{equation}
	V_\mathrm{y,los} = L \times {2\sin(\sfrac{{\theta}_\mathrm{y}}{2})}/{t}
	\label{eqn:vsrcvert}
\end{equation}
\end{linenomath*}

\medskip

\noindent where $L$ is the distance, ${\theta}_{x} = \sqrt{{\theta}^{2}-{\delta}E^{2}}$ and ${\theta}_{y} = {\delta}E$ are the $x$ and $y$ components of the angle through which the radio source moves, ${\theta}$, and ${\delta}E$ is the change in elevation.  

The ionospheric velocities measured by radar have been translated into magnitudes, $V_{ion}$, with directions given in terms of azimuth measured eastwards from north, $A_{ion}$.  The x and y components of ionospheric velocity can then be calculated with relation to the azimuth, $A_{los}$, and elevation, $E$, of the line of sight:

\begin{linenomath*}
\begin{equation}
	{\delta}A = A_\mathrm{ion} - A_\mathrm{los}
	\label{eqn:deltaa}
\end{equation}
\end{linenomath*}

\begin{linenomath*}
\begin{equation}
	V_\mathrm{x,ion} = V_\mathrm{ion}\sin({\delta}A)
	\label{eqn:vionhoriz}
\end{equation}
\end{linenomath*}

\begin{linenomath*}
\begin{equation}
	V_\mathrm{y,ion} = V_\mathrm{ion}\cos({\delta}A)\sin(E)
	\label{eqn:vionvert}
\end{equation}
\end{linenomath*}

\medskip

The total transverse velocity can then be stated as:

\begin{equation}
	V^{2} = {(V_\mathrm{x,ion}-V_\mathrm{x,los})}^{2}+(V_\mathrm{y,ion}-V_\mathrm{y,los})^{2}
	\label{eqn:perpvel}
\end{equation}

\medskip

Taking the arc curvature model as given in Section~\ref{sec:arcfit} ($L=2CV^{2}$ where $C$ is the arc curvature) and combining with Equations \ref{eqn:vsrchoriz} to \ref{eqn:perpvel}, a quadratic equation in terms of $L$ can be formed:

\begin{equation}
\begin{split}
	8C\left(\left(\frac{\sin(\sfrac{{\theta}_\mathrm{x}}{2})}{t}\right)^{2}+\left(\frac{\sin(\sfrac{{\theta}_\mathrm{y}}{2})}{t}\right)^{2}\right)L^{2} \\ 
	-8CV_\mathrm{ion}\sin({\delta}A) \left(\frac{\sin(\sfrac{{\theta}_\mathrm{x}}{2})}{t}\right)L & \\ 
	-8CV_\mathrm{ion}\cos({\delta}A)\sin(E)\left(\frac{sin(\sfrac{{\theta}_\mathrm{y}}{2})}{t}\right)L & \\ 
	-L & \\ 
	+2CV_\mathrm{ion}^{2}\left(\sin^{2}({\delta}A)+\cos^{2}({\delta}A)\sin^{2}(E)\right) & = 0
\end{split}
\label{eqn:Lquadratic}
\end{equation}

\medskip

\begin{figure}
	\centering
	\includegraphics[width=9cm]{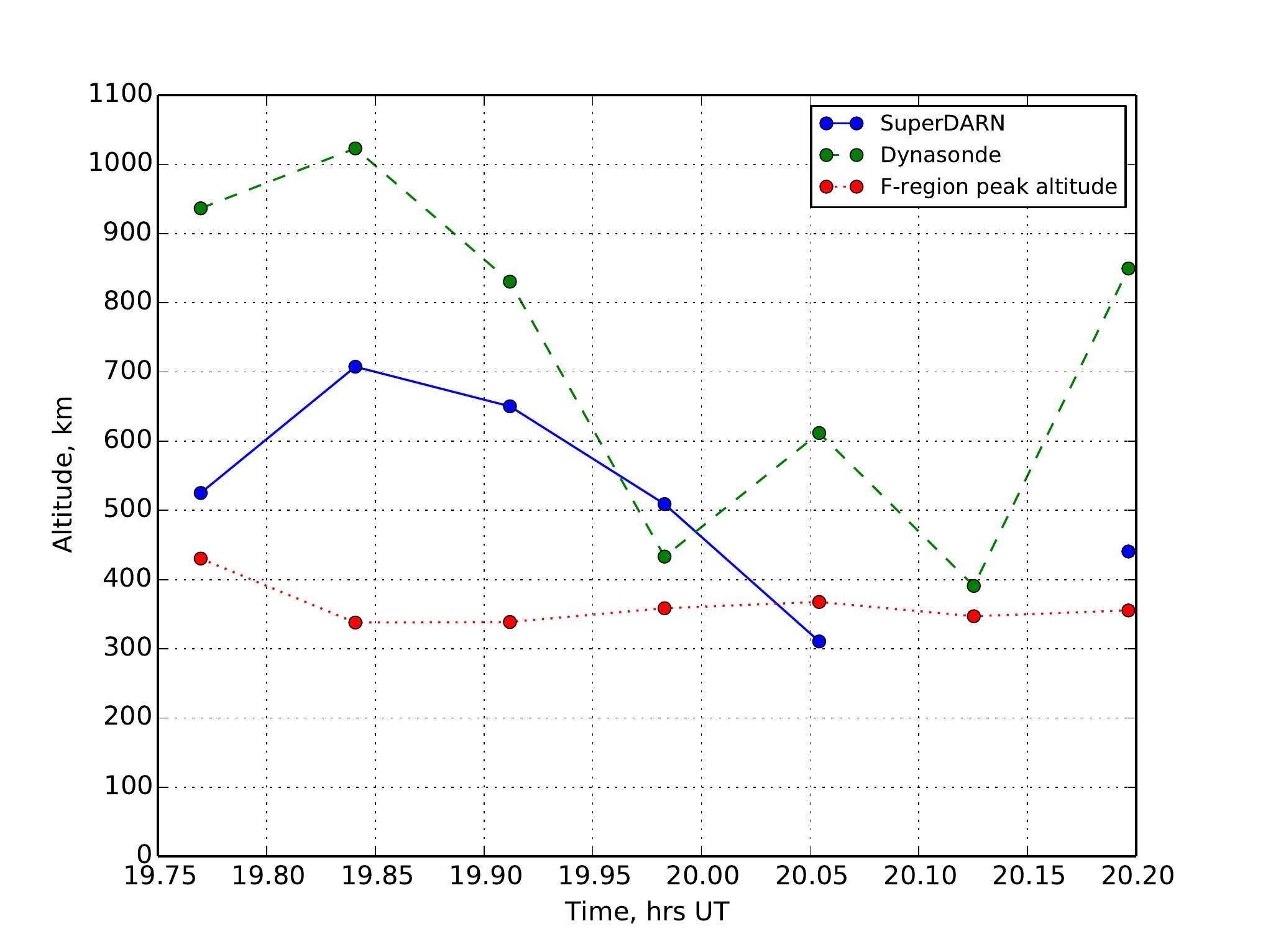}
	\caption{Dominant scattering altitudes estimated by combining the scintillation arc model with perpendicular velocities measured by the two radar systems.  Also shown are the F-region peak altitudes measured by the Troms\o~dynasonde.} 
	\label{fig:scatheights}
\end{figure}

Equation \ref{eqn:Lquadratic} is solved using the standard quadratic equation and distances calculated for each block using both SuperDARN and dynasonde velocities.  Using the negative root of the standard quadratic equation results in distances of $\sim$~0, so distances were calculated using only the positive root.  The calculated distances have been converted into altitudes and plotted in Figure~\ref{fig:scatheights}.

It can be seen that the altitudes calculated from the SuperDARN velocities are within the F-region, but most are significantly higher than the peak height measured by the Troms\o~dynasonde.  There was no real-valued solution to the quadratic equation for one of the SuperDARN velocity points.  The dynasonde velocities indicate rather higher altitudes in general.

\subsection{Estimating Velocity}
\label{sec:scatvel}

The altitudes of peak F-region density measured by the Troms\o~dynasonde can be used themselves as an estimate of the dominant altitudes of the scattering and so combined with the scintillation arc modelling to estimate a range of possible ionospheric velocities perpendicular to the line of sight.  

\begin{figure}
	\centering
	\includegraphics[width=9cm]{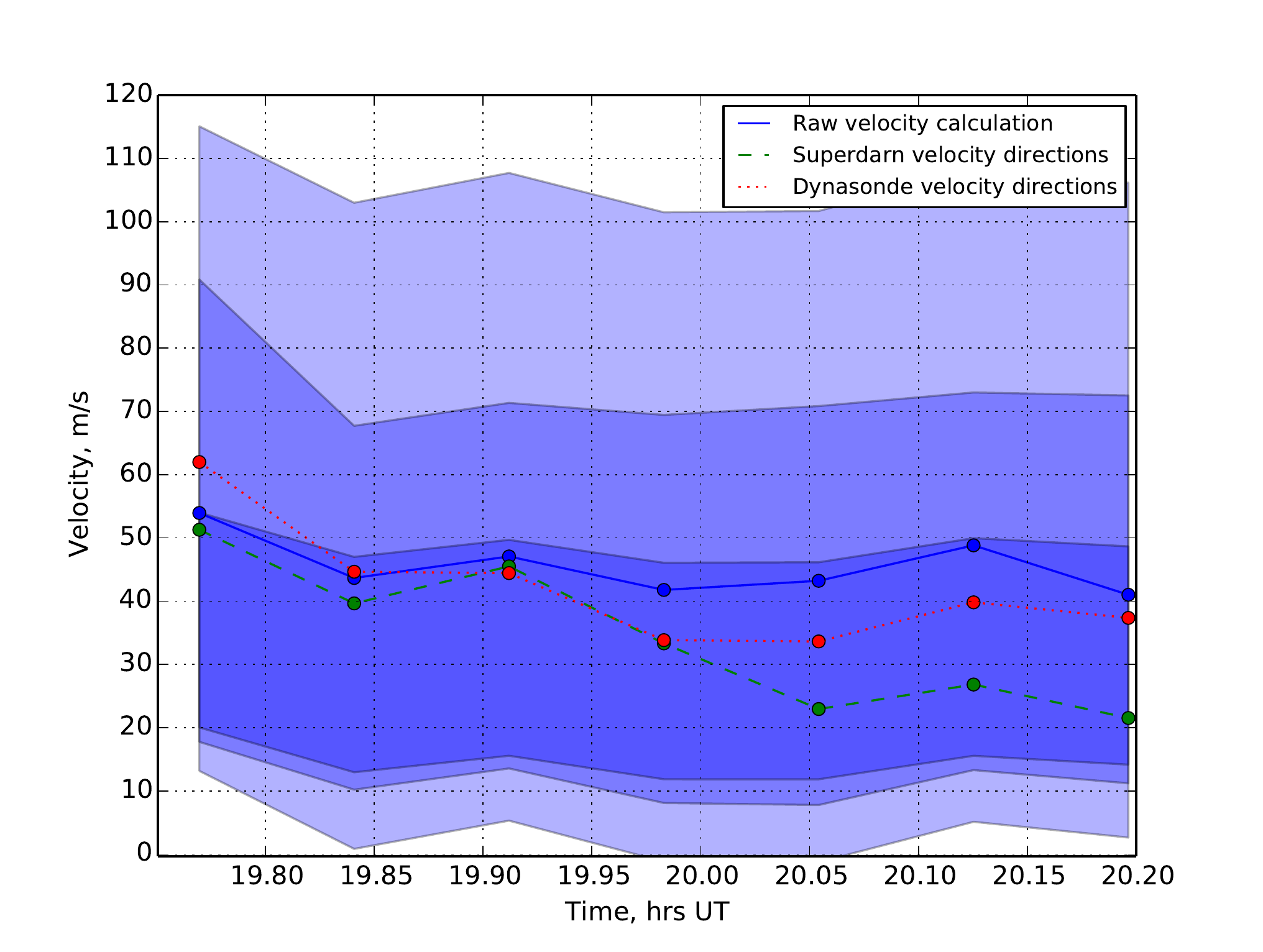}
	\caption{Possible values for the ionospheric velocity component perpendicular to the line of sight.  The shaded blue regions represent the possible range of velocities calculated assuming scattering from 200\,km altitude (narrowest range with the darkest blue), scattering from the F-layer peak as measured by the Troms\o~dynasonde, and scattering from 600\,km altitude (widest range with the lightest blue).  Also shown are three lines: the blue solid line represents the direct calculation of the overall velocity using the scintillation arc curvatures and F-layer peak altitudes; the green dashed line represents velocities from the same calculation, but assuming that the directions of ionospheric velocity are equal to those measured by SuperDARN; the red dotted line represents velocities assume directions of ionospheric velocity equal to those measured by the dynasonde. }
	\label{fig:minmaxvel}
\end{figure}

The basic $L=2CV^{2}$ is used to calculate velocities in this instance but, as described in section \ref{sec:scatheight}, this velocity contains components from both the ionosphere and the movement of the line of sight to the radio source through it.  The latter can be calculated but, since these components are vectors (as detailed in equation \ref{eqn:perpvel}), both the speed and direction of the ionospheric component would need to be estimated.  However, it is not possible to obtain both pieces of information from this simple model; equation \ref{eqn:Lquadratic} can be re-arranged in terms of the magnitude $V_{ion}$ or the azimuth direction ${\delta}A$ but it is not possible to calculate both.  Therefore we can only look at the magnitudes and estimate a range of possible speeds.

The magnitude of the line of sight velocity, $V_{los}$ is calculated from equations \ref{eqn:vsrchoriz} and \ref{eqn:vsrcvert} as $V_{los} = \sqrt{V_{x,los}^{2}+V_{y,los}^{2}}$.  Given the unknown direction of the ionospheric velocity relative to this, adding and subtracting this number from the modelled $V$ will give the maximum and minimum possible values for $V_{ion}$ respectively.  The results of this exercise are given in Figure \ref{fig:minmaxvel}. 

In this figure the calculated overall velocity $V$ is plotted as a blue solid line. Also plotted are $V_{ion}$ velocities calculated assuming the velocity directions given by the SuperDARN and Dynasonde data, using equation \ref{eqn:Lquadratic} re-arranged in terms of $V_{ion}$ and solved for the positive root using the standard quadratic equation.  Using the negative root resulted in negative velocities whereas this parameter, as derived here, is a magnitude.  Three possible ranges of $V_{ion}$ are shown, calculated assuming three different sets of altitudes.  From the tomography density plots, we estimate a minimum and maximum likely altitude for the scattering as 200\,km and 600\,km respectively.  These values represent the altitudes at which the density has dropped to approximately half of its peak value.  In Figure \ref{fig:minmaxvel}, the narrowest range (shown in the darkest blue) was calculated assuming scattering from the lowest altitude and the widest range (shown in the lightest blue) was calculated assuming scattering from the highest altitude.  The middle range was calculated assuming altitudes equal to the F-layer peak height measured by the Troms\o~dynasonde. 




\subsection{Increased Time Resolution}
\label{sec:timeres}

A number of the arcs in blocks 10-16 show some broadening which could be interpreted as the superposition of two arcs with slightly different curvatures.  Such a scenario could indicate scattering from different altitudes and/or with different velocities, or it can indicate that the 256\,s block length averages over slight changes in conditions.  There is little evidence for the former scenario in this case, leaving the latter to be considered.

\begin{figure}
	\centering
	\includegraphics[width=9cm]{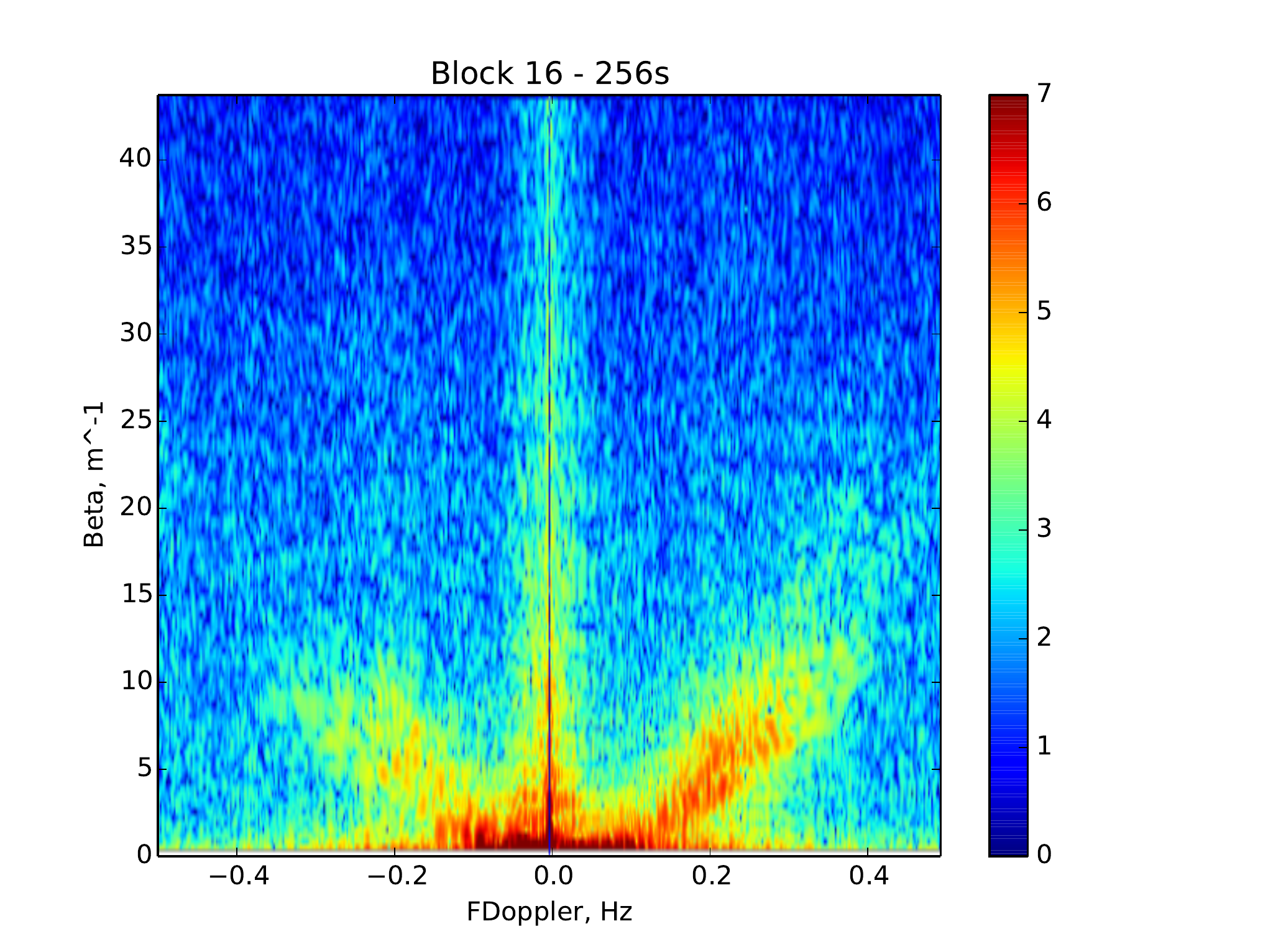}
	\includegraphics[width=9cm]{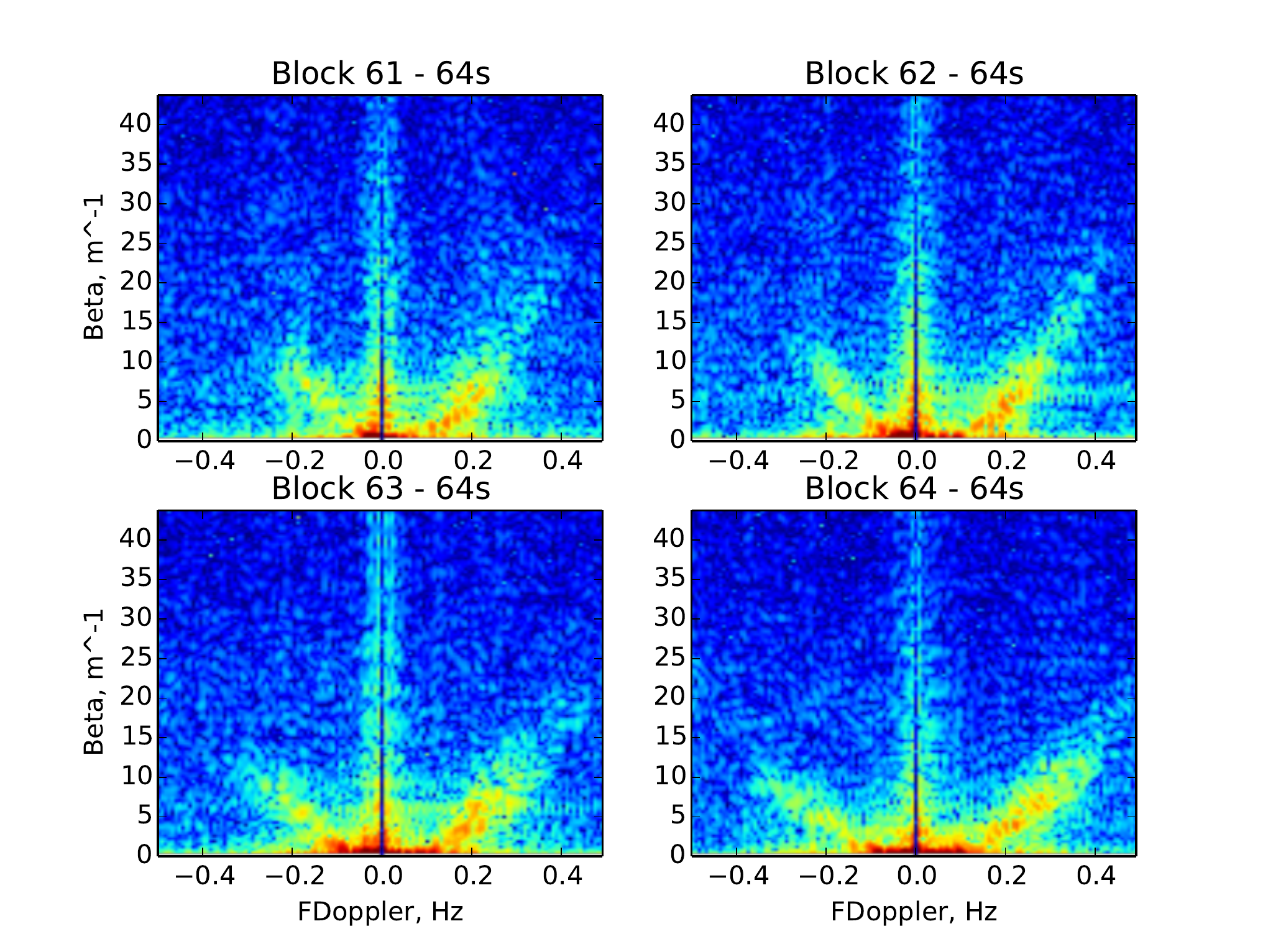}
	\caption{Secondary spectra for Block 16.  The scale is logarithmic with units equivalent to dB/10. Top: 256\,s block length; Bottom: Corresponding spectra for 4x 64\,s block length, using the same scale as the top plot.}
	\label{fig:arcs64s}
\end{figure}

To do so, the block length was reduced as low as 64\,s.  Although this reduces the signal-to-noise ratio of the secondary spectra, the scintillation arcs are still readily apparent and do indeed appear narrower and more clearly defined.  This is illustrated in Figure~\ref{fig:arcs64s} which presents a comparison between the original secondary spectrum for block 16 and spectra for the individual 64\,s blocks which comprise it.

The comparison demonstrates clearly that the broadening of the scintillation arcs is due to changing conditions through a single 256\,s block and not a simultaneous superposition of scattering regimes with different altitudes and/or velocities.

\begin{figure}
	\centering
	\includegraphics[width=9cm]{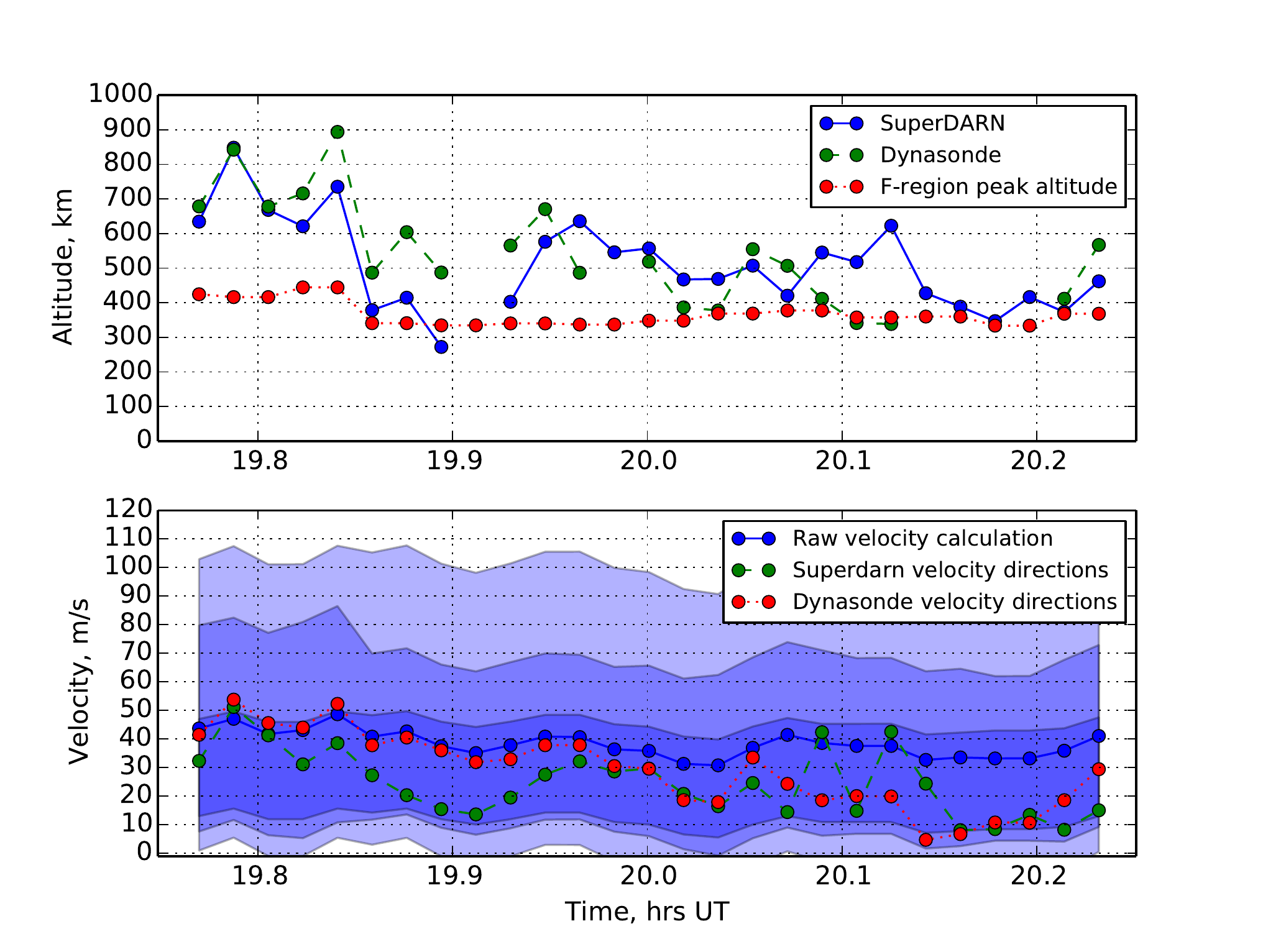}
	\caption{Top: Dominant scattering altitudes estimated by combining the scintillation arc model with perpendicular velocities measured by the two radar systems.  Also shown are the F-region peak altitudes measured by the Troms\o~dynasonde. . Bottom: Possible values for the ionospheric velocity component perpendicular to the line of sight.  The shaded blue regions represent the possible range of velocities calculated assuming scattering from 200\,km altitude (narrowest range with the darkest blue), scattering from the F-layer peak as measured by the Troms\o~dynasonde, and scattering from 600\,km altitude (widest range with the lightest blue).  Also shown are three lines: the blue solid line represents the direct calculation of the overall velocity using the scintillation arc curvatures and F-layer peak altitudes; the green dashed line represents velocities from the same calculation, but assuming that the directions of ionospheric velocity are equal to those measured by SuperDARN; the red dotted line represents velocities assume directions of ionospheric velocity equal to those measured by the dynasonde.}
	\label{fig:scatdetails64s}
\end{figure}

The calculated scattering altitudes and possible velocity range for the 64\,s blocks are given in Figure~\ref{fig:scatdetails64s}.  These were calculated using the original SuperDARN data (i.e., before the averaging of every four points for the 256\,s blocks) and dynasonde data padded to account for the 2-minute cadence of the original data. 

The scattering altitudes show some substantial variation within the F-region, only sometimes becoming as low as the altitude of the peak of the F-region, suggesting that the scattering is more dominated by the topside ionosphere.  The velocity ranges shown are as described in section \ref{sec:scatvel}.  Although the overall range of possible velocities is quite high, they do confirm that the velocity is most likely to be well under 100\,m\,s$^{-1}$.

\section{Conclusions}

We present the first multi-octave low frequency observations of ionospheric scintillation, using the KAIRA station.  These show in dynamic spectra the evolution of the scattering regime from weak scatter at the highest observing frequencies to strong scatter and the effects of refraction by large-scale density structures in the ionosphere at the lowest frequencies observed.

The secondary spectra, as calculated from the 2-D FFT of the dynamic spectra, often show an arc structure similar to the ``scintillation arcs'' noted in observations of the interstellar scintillation of pulsars.  Here, we apply the basic scattering theory developed for this latter case to our observations of ionospheric scintillation to estimate velcocity perpendicular to the line of sight to the radio source and the altitude(s) dominating the scattering.

Since these parameters are linked in the modelling, it is necessary to use other observations to provide estimates of one so that the other may be calculated.  In this case, velocities measured by SuperDARN and the Troms\o~dynasonde are used.  However, neither of these are exactly coincident with the region being probed by the scintillation observations.  Nevertheless scattering altitudes consistent with the F-region are found when modelling the scintillation arcs using the horizontal ionospheric velocities measured by SuperDARN and the Troms\o ~dynasonde.  In some cases these match the peak altitude of the F-region, as measured by the dynasonde.  However, in most cases the scattering altitudes appear higher than this peak, indicating that the scattering may be dominated more by the topside ionosphere.

Estimating ionospheric velocities using this technique is more complex as it involves the estimation of two parameters, the magnitude and direction, which are, again, inherently linked in the modelling and require knowledge of one for the other to be calculated.  Nevertheless, a range of possible velocity magnitudes can be calculated given the scattering altitude.  The range is, however, broad and in the results presented here can only confirm that the velocity is most likely to be below 100\,m\,s$^{-1}$. 


In this paper we give only an example of the analyses which can be performed with broadband data such as these.  While we have been able to use a simple model here to obtain estimates of altitudes dominating the scattering and ionospheric velocities, it is clear that more information could be obtained if a full scattering model is applied.  This will be the subject of future work.

With the densely-packed core of stations in the center of the LOFAR array it is possible to do more.  Observations already taken using the LOFAR stations demonstrate that ionospheric scintillation is readily apparent at these frequencies, even at mid-latitudes.  With the main LOFAR core, it is possible to use temporal cross-correlation between time series' recorded by indivdual stations to estimate ionospheric velocities to feed into modelling of scintillation arcs seen in the secondary spectra.  Data have already been taken and will be the subject of future work.

Also of particular interest is the work of \citet{Briskenetal:2010} in which radio astronomy imaging techniques have been combined with secondary spectrum analyses to measure accurately the locations of an image of a pulsar, scattered by the interstellar medium.  Applying this technique to observations using the LOFAR core offers the prospect of ``imaging'' ionospheric scintillation for the first time. 

\begin{acknowledgements}
KAIRA was funded by the University of Oulu and the FP7 European Regional Development Fund and is operated by Sodankyl\"a Geophysical Observatory.  This work has been funded in part by the Academy of Finland (application number 250252, Measurement Techniques for Multi-Static Incoherent Scatter Radars and application number 250215, Finnish Programme for Centre of Excellence in Research 2012- 2017).  This paper is based on results obtained with LOFAR equipment. LOFAR \citep{LOFAR-reference-paper:2013} is the Low Frequency Array designed and constructed by ASTRON.  Unless otherwise specified, the data used in this paper will be made available upon email request to R.A.Fallows: fallows@astron.nl or rafallows@gmail.com.  KAIRA data policy states that any  project making non-trivial use of KAIRA data is required to have a Sodankyl\"a Geophysical Observatory staff member as co-author on published papers unless agreed otherwise.
\end{acknowledgements}


\begin{thebibliography}{19}
\expandafter\ifx\csname natexlab\endcsname\relax\def\natexlab#1{#1}\fi

\bibitem[{{\it {Brisken} et~al.\/}(2010){\it {Brisken}, {Macquart}, {Gao},
  {Rickett}, {Coles}, {Deller}, {Tingay}, and {West}\/}}]{Briskenetal:2010}
{Brisken}, W.~F., J.-P. {Macquart}, J.~J. {Gao}, B.~J. {Rickett}, W.~A.
  {Coles}, A.~T. {Deller}, S.~J. {Tingay}, and C.~J. {West}, {100 {$\mu$}as
  Resolution VLBI Imaging of Anisotropic Interstellar Scattering Toward Pulsar
  B0834+06}, {\it Astrophysical Journal\/}, {\it 708\/}, 232--243, 2010.

\bibitem[{{\it Cannon et~al.\/}(2006){\it Cannon, Groves, Fraser, Donnelly, and
  Perrier\/}}]{Cannonetal:2006}
Cannon, P.~S., K.~Groves, D.~J. Fraser, W.~J. Donnelly, and K.~Perrier, Signal
  distortion on vhf/uhf transionospheric paths: First results from the wideband
  ionospheric distortion experiment, {\it Radio science\/}, {\it 41\/}, 2006.

\bibitem[{{\it {Chisham} et~al.\/}(2007)}]{Chishametal:2007}
{Chisham}, G., et~al., {A decade of the Super Dual Auroral Radar Network
  (SuperDARN): scientific achievements, new techniques and future directions},
  {\it Surveys in Geophysics\/}, {\it 28\/}, 33--109, 2007.

\bibitem[{{\it Cordes et~al.\/}(2006){\it Cordes, Rickett, Stinebring, and
  Coles\/}}]{Cordesetal:2006}
Cordes, J.~M., B.~J. Rickett, D.~R. Stinebring, and W.~A. Coles, Theory of
  parabolic arcs in interstellar scintillation spectra, {\it The Astrophysical
  Journal\/}, {\it 637\/}, 346, 2006.

\bibitem[{{\it Gillies et~al.\/}(2009){\it Gillies, Hussey, Sofko, McWilliams,
  Fiori, Ponomarenko, and St-Maurice\/}}]{Gilliesetal:2009}
Gillies, R., G.~Hussey, G.~Sofko, K.~McWilliams, R.~Fiori, P.~Ponomarenko, and
  J.-P. St-Maurice, Improvement of superdarn velocity measurements by
  estimating the index of refraction in the scattering region using
  interferometry, {\it Journal of Geophysical Research: Space Physics
  (1978--2012)\/}, {\it 114\/}, 2009.

\bibitem[{{\it {Greenwald} et~al.\/}(1995)}]{Greenwaldetal:1995}
{Greenwald}, R.~A., et~al., {Darn/Superdarn: A Global View of the Dynamics of
  High-Lattitude Convection}, {\it Space Science Reviews\/}, {\it 71\/},
  761--796, 1995.

\bibitem[{{\it Jenkins and Watts\/}(1969)}]{JenkinsWatts:1969}
Jenkins, G., and D.~Watts, Spectral analysis and its applications. emerson,
  1969.

\bibitem[{{\it Knepp and Nickisch\/}(2009)}]{KneppandNickisch:2009}
Knepp, D.~L., and L.~Nickisch, Multiple phase screen calculation of wide
  bandwidth propagation, {\it Radio Science\/}, {\it 44\/}, 2009.

\bibitem[{{\it Lecacheux et~al.\/}(2004){\it Lecacheux, Konovalenko, and
  Rucker\/}}]{Lecacheuxetal:2004}
Lecacheux, A., A.~Konovalenko, and H.~Rucker, Using large radio telescopes at
  decametre wavelengths, {\it Planetary and Space Science\/}, {\it 52\/},
  1357--1374, 2004.

\bibitem[{{\it Markkanen et~al.\/}(1995){\it Markkanen, Lehtinen, Nygren,
  Pirttila{\`E}, Henelius, Vilenius, Tereshchenko, and
  Khudukon\/}}]{Markkanenetal:1995}
Markkanen, M., M.~Lehtinen, T.~Nygren, J.~Pirttila{\`E}, P.~Henelius,
  E.~Vilenius, E.~Tereshchenko, and B.~Khudukon, Bayesian approach to satellite
  radiotomography with applications in the scandinavian sector, in {\it Annales
  geophysicae\/}, vol.~13, pp. 1277--1287, Copernicus, 1995.

\bibitem[{{\it McKay-Bukowski et~al.\/}(2014)}]{KAIRA-reference-paper:2014}
McKay-Bukowski, D., et~al., {KAIRA}: the {K}ilpisja\"rvi {A}tmospheric
  {I}maging {R}eceiver {A}rray -- system overview and first results, {\it
  accepted for publication in IEEE Transactions on Geoscience and Remote
  Sensing\/}, 2014.

\bibitem[{{\it Meyer-Vernet et~al.\/}(1981){\it Meyer-Vernet, Daigne, and
  Lecacheux\/}}]{MeyerVernetetal:1981}
Meyer-Vernet, N., G.~Daigne, and A.~Lecacheux, Dynamic spectra of some
  terrestrial ionospheric effects at decametric wavelengths-applications in
  other astrophysical contexts, {\it Astronomy and Astrophysics\/}, {\it 96\/},
  296--301, 1981.

\bibitem[{{\it Narayan\/}(1992)}]{Narayan:1992}
Narayan, R., The physics of pulsar scintillation, {\it Philosophical
  Transactions of the Royal Society of London. Series A: Physical and
  Engineering Sciences\/}, {\it 341\/}, 151--165, 1992.

\bibitem[{{\it Nickisch\/}(1992)}]{Nickisch:1992}
Nickisch, L., Non-uniform motion and extended media effects on the mutual
  coherence function: An analytic solution for spaced frequency, position, and
  time, {\it Radio science\/}, {\it 27\/}, 9--22, 1992.

\bibitem[{{\it {Nygr{\'e}n} et~al.\/}(1997){\it {Nygr{\'e}n}, {Markkanen},
  {Lehtinen}, {Tereshchenko}, and {Khudukon}\/}}]{Nygrenetal:1997}
{Nygr{\'e}n}, T., M.~{Markkanen}, M.~{Lehtinen}, E.~D. {Tereshchenko}, and
  B.~Z. {Khudukon}, {Stochastic inversion in ionospheric radiotomography}, {\it
  Radio Science\/}, {\it 32\/}, 2359--2372, 1997.

\bibitem[{{\it Rietveld et~al.\/}(2008){\it Rietveld, Wright, Zabotin, and
  Pitteway\/}}]{Rietveldetal:2008}
Rietveld, M., J.~Wright, N.~Zabotin, and M.~Pitteway, The {T}roms{\o}
  dynasonde, {\it Polar Science\/}, {\it 2\/}, 55--71, 2008.

\bibitem[{{\it Rishbeth and Williams\/}(1985)}]{RishbethWilliams:1985}
Rishbeth, H., and P.~Williams, The eiscat ionospheric radar: the system and its
  early results, {\it Q. Jl. R. Astr. Soc.\/}, {\it 26\/}, 478--512, 1985.

\bibitem[{{\it Stinebring et~al.\/}(2001){\it Stinebring, McLaughlin, Cordes,
  Becker, Espinoza~Goodman, Kramer, Sheckard, and
  Smith\/}}]{Stinebringetal:2001}
Stinebring, D., M.~McLaughlin, J.~Cordes, K.~Becker, J.~Espinoza~Goodman,
  M.~Kramer, J.~Sheckard, and C.~Smith, Faint scattering around pulsars:
  Probing the interstellar medium on solar system size scales, {\it Astrophys.
  J. Letts.\/}, {\it 549\/}, L97, 2001.

\bibitem[{{\it {van Haarlem} et~al.\/}(2013)}]{LOFAR-reference-paper:2013}
{van Haarlem}, M.~P., et~al., {LOFAR: The LOw-Frequency ARray}, {\it \aap\/},
  {\it 556\/}, A2, 2013.

\end{thebibliography}

\end{article}
\end{document}